# Conditional Flow Matching for Generation of 3D Multi-variable Instantaneous Urban Microclimate Fields

Peng Liu[1], Shaoxiang Qin[1,2], Theodore Potsis[1], Lili Ji[1], Dingyang Geng[1], and Liangzhu Leon Wang[1,*]

[1]Concordia University, Centre for Zero Energy Building Studies, Department of Building, Civil and Environmental Engineering, Montreal H3G 1M8, Canada

[2]McGill University, School of Computer Science, Montreal, H3A 0E9, Canada

# Abstract

Rapid and accurate prediction of urban wind and temperature fields is important for urban microclimate design and climate adaptation. Large-eddy simulation (LES) effectively resolves these instantaneous fields, but its application is limited in iterative design of urban microclimate applications due to high computational cost. Existing regressive data-driven models offers quick outputs, but they produce only deterministic point predictions that inherently fail to represent turbulent stochasticity. This paper adopts a novel generative framework of Conditional Flow Matching (CFM) that uses building geometry and mean flow as guidance to generate plausible three-dimensional instantaneous velocity and temperature fields for urban microclimate in seconds. To overcome the GPU memory bottleneck of pixel space 3D generation, the model operates in parallel on overlapping pixel space through a shared-noise initialization that preserves high spatial continuity of flow structure across the entire domain. Against reference LES data, the CFM surrogate can rapidly and accurately restore the first-order statistics with Normalized Root Mean Square Error (NRMSE) of 2.99% for wind and 1.77% for temperature, second-order turbulence metrics with NRMSE of 7.17% for wind and 8.84% for temperature, turbulent kinetic energy with NRMSE of 7%, probability density function and vertical profiles in representative locations. Wind engineering application of local gust prediction demonstrate that the speed and accuracy of CFM, supporting the use of generative AI for making turbulence-aware resilient urban design and climate adaptation more computationally feasible.

**Highlights**

- Conditional flow matching generates stochastic 3D urban wind and temperature fields.
- Geometry and mean flow alone guide snapshot generation in about 10 seconds.
- Shared-noise patches preserve spatial continuity across a 1.2 × 1.2 km domain.
- Mean and standard-deviation fields match LES with NRMSEs below 3% and 9%.
- Local gust estimates agree with reference simulations under quasi-steady assumptions.

# 1. Introduction

With the global urban population projected to reach 70% by 2050, the 21st century is characterized by an unprecedented expansion of the artificial land cover (Toparlar et. al, 2017). This rapid urbanization is not merely a demographic shift but also a systematic change in built environment, marked by the growing density and verticality of urban areas. As an outcome, fundamentally altered urban aerodynamics and energy transfer processes are observed due to increased flow heterogeneity, turbulent mixing, and changes in the convective heat transfer. This alteration triggers a series of microclimatic challenges for urban residency, industrial production, and activities. Most notably is the intensification of the Urban Heat Island (UHI) effect, which is presented by higher air and land surface temperature in built environment, are severely affecting the residential comfortability and even lives. Nowadays, sustainable, climate-sensitive, and occupant-oriented urban planning requires a granular understanding of how these new "synthetic" topographies interact with the urban boundary layer.

Urban microclimate governs a range of phenomena that directly affect the quality of life in cities, from pedestrian wind comfort and natural ventilation to pollutant dispersion, building aerodynamics, and outdoor thermal comfort (Yang et al., 2023). Understanding these phenomena requires methods capable of capturing instantaneous, transient turbulent structures, not restricted to mean flow field or steady state. Pedestrian comfort assessments depend on the probability of encountering wind gusts above a threshold, ventilation studies require recirculation flow prediction in complex flow structures, and safety analyses demand knowledge of peak wind loads that only appear in the tails of the velocity distribution (Blocken and Carmeliet, 2004; Hågbo et al., 2022). For all these applications, the critical information lies in the stochastic variability of the wind field, not the mean.

Computational Fluid Dynamics (CFD) approaches like Reynolds Averaged Navier Stokes (RANS) and Large Eddy Simulation (LES) are the most capable tools to model urban microclimates. Regarding RANS, the approximation of fluctuations that are needed for more detailed studies is insufficient, and in combination with the increasing computational resources a transition to LES has been seen in the current state-of-the-art (Potsis et al., 2023). LES resolves the flow field that carry most of high frequency and stochastic fluctuations and has been applied successfully to realistic urban geometries (Tolias et al., 2018; Letzel et al., 2008; Cheng and Yang, 2022). However, besides High-Performance Computing accessibility, LES remains computationally prohibitive for real-time iterative design workflows: a single urban scenario on a domain of several hundred million grid cells requires hours to days of CPU time and producing the necessary for probabilistic analysis multiplies this cost further (Blocken, 2018; Yang, 2015).

This computational bottleneck has motivated the work of data-driven surrogates that approximate CFD outputs with remarkably reduced computational costs. Deterministic or regressive deep learning models, such as Convolutional Neural Networks (CNN), Fourier Neural Operators (FNO), Graph Neural Networks (GNN) and others, have demonstrated a series of success in predicting time-averaged or steady-states velocity and temperature fields (Bhatnagar et al., 2019; Liu et al., 2023; Peng et al., 2024; Qin et al., 2025). However, these regressive frameworks inherently struggle to capture the transient, multi-scale nature of turbulent flows. Because they are typically optimized to minimize a point-to-point loss, they tend to predict the conditional expectation, effectively "averaging out" the stochastic fluctuations inherent in turbulence (Wang et al., 2025; Calzolari and Liu, 2021). While some recent deterministic data-driven surrogates can predict scalar turbulence statistics such as turbulence intensity, they remain unable to provide the full realization of physically plausible instantaneous flow states that urban applications demand (Duraisamy et al., 2018; Veiga-Piñeiro et al., 2025).

Generative models offer a promising path to tackle such limitations for these regressive neural networks, such as providing manifolds of turbulent uncertainty (Du et al., 2024). Unlike deterministic regression, probabilistic generative models are designed to learn the underlying data distribution, which benefit the tasking aiming to capture the stochastic nature of turbulence (Li et al., 2023). This allows them to sample from a manifold of physically plausible states, capturing the transient variations that regressive models inherently suppress or ignore, while such variability from norm affects the reliability of the urban design and comfortability of residents (Bukka et al., 2021). Generative Adversarial Networks (GANs) have been applied to instantaneous urban wind fields to recover instantaneous airflow from POD-LSE and also found to improve the wake region and near-building wind prediction by embedding frequency features (Kastner and Dogan, 2023; Goodfellow et al., 2014; Hu et al., 2023; Wang et al., 2024). But the training of CGAN has proven unstable due to the adversarial minimax optimization between the generator and discriminator, particularly for high-dimensional 3D vector fields. Variational Autoencoders (VAEs) provide fast sampling and stable training and has been adopted to generate and reconstruct flow field quantities across engines and airfoils (Posch et al., 2022; Wang et al., 2021), but their KL-regularization often produces over-smoothed or irregular dynamics that fail to capture sharp turbulent structures (Kingma and Welling, 2013; Eivazi et al., 2022; (Dubois et al., 2022). Denoising Diffusion Probabilistic Models (DDPMs) achieve higher sampling quality for reconstructing fine flow field but suffer from slow training and inference due to the requirement of hundreds of sampling steps that are calculated using Stochastic Differential Equation (SDE) (Song et al., 2020; Ho et al., 2020; Gao et al., 2023; Tahmasebi et al., 2025). Conditional Flow Matching (CFM) has recently emerged as a mathematically robust and effective alternative that leverages the strengths of former generative models while avoiding their principal weaknesses (Lipman et al., 2023; Albergo and Vanden-Eijnden, 2022). CFM adopts Ordinary Differential Equation (ODE) for calculating sampling timesteps, along with Optimal Transport (OT) to learn a retractable, direct, and highly efficient intermediate vector field that will lead the sample points from the source gaussian noise to the target data distribution (Tong et al., 2023; Tong et al., 2024). Compared to Diffusion models, with ODE solution, CFM produce retraceable, deterministic and more straight-forward transport paths for sample points that require less integration steps (usually 1 to 20, compared to 100+ for Diffusion); Compared to GANs, it avoids adversarial training entirely and provide stable, efficient, and simulation-free training, which is highly valued in training a dataset of high-dimensional vectors; Unlike traditional VAEs, CFM avoids blurry results by indirectly learning the intermediate vector fields instead of directly operating on pixel-level regression. It resolves three key VAE flaws: tendency of regression to the mean, information-losing latent bottlenecks, and posterior collapse. By learning straight-line trajectories, CFM preserves high-frequency turbulent details that VAEs might smooth out. Recent work has demonstrated the potential of CFM for fluid flow applications, such as near-wall turbulence generation with uncertainty quantification (Parikh et al., 2025), and full-scale aircraft simulations on unstructured meshes (Ramos et al., 2026).

In generation tasks for large 3D urban instantaneous CFD flow field, there exist a long theoretical conflict between pixel-space direction generation with computational resource limitations, especially GPU ram for training (Du et al., 2024; Qin et al., 2025). Generative models for high-resolution and spatial-temporal data commonly generalize uses latent space training (such as using a pretrained VAE to compress the full domain for the training of diffusion) (Rombach et al., 2022; Vahdat et al., 2021). Such frameworks can fairly reduce the computational costs of a 3D mesh, but due to the compression bottleneck between encoding and decoding, high-frequency local structures, such as high fluctuations, separation around sharp building edges will be lost and create blurred artifacts (Vahdat et al., 2021). Comparing latent space training and inference, generation directly on pixel-space retains high level of spatial fidelity, but its practicality is heavily limited by computational resources on a single GPU for large urban domain (Qin et al., 2025).

To tackle such trade-off, recently papers in the machine learning field for urban microclimate studies has resorted to use overlapping patches to prevent memory outage and maintain boundary continuity for global flow field (Arakawa et al., 2023; Lin et al., 2024). Meanwhile, give the advantages of CFM model being a novel alternative generative paradigm as opposed to GANs and Diffusion, the field of fluid mechanics took the advantages of efficient trajectory sampling and agreement to the conditional guidance for more generative tasks (Lipman et al., 2023; Lipman et al., 2024; Tong et al., 2024). However, existing CFM studies have focused on relatively small-scale applications and have not yet been demonstrated for large, spatially heterogeneous 3D urban LES flow fields (Du et al., 2024; Ramos et al., 2026). A vital gap therefore remains in developing scalable methods that can generate large transient wind-field ensembles while preserving spatial fidelity and global consistency (Arakawa et al., 2023).

In this paper, we develop a CFM-based generative surrogate that produces unlimited number of high-fidelity three-dimensional instantaneous urban microclimate snapshots of velocity components (u, v, w) and temperature (T), conditioned on building geometry and a steady-state mean flow field. The novel CFM framework presented in the paper for urban LES modelling has three major objectives:

1. Take the advantage of the efficient CFM generative paradigm and successfully apply it to the real urban microclimate scenario.

2. Overcome the common GPU RAM deficiency for probabilistic generative models when operating in large 3-dimensional tensor space with multi-variable inputs, contributing towards scalable methods.

3. Use steady-state flow field as well as urban geometry as guiding conditional inputs to generate instantaneous urban microclimate snapshots. Such practice bridged the gap for many studies in the state-of-the-art that predict the mean flow field by offering a possibility to expand their works into plausible and dynamic transient states ensembles that is valuable for extreme-case analysis and real-world problems.

The structure of the paper is as follows: first we present the methodology of this work in Section 2, by discussing the LES simulation data, then the CFM framework and the patch-wise generation of snapshots. In Section 3 the main validation results of the paper are presented for instantaneous flow generation, mean flow and turbulence. Also, comparisons of PDFs and vertical profiles in characteristic locations are presented. Next in Section 4 we further the discussion of CFM for predicting wind as a practical implementation and present some potential future target of this work. Finally, we conclude the papers with its main findings in Section 5.

# 2. Methodology

## 2.1 Numerical simulations

The training data for the generative model are produced using CityFFD, a GPU-accelerated CFD solver that addresses the computational bottleneck of traditional methods by employing a semi-Lagrangian approach coupled with a fractional step method (Katal et al., 2019; Mortezazadeh et al., 2021; Yang et al., 2022; Mortezazadeh et al., 2022). CityFFD implements a high-order backward-forward sweep interpolation scheme to mitigate the numerical dissipation typically associated with semi-Lagrangian advection, ensuring that high-frequency flow features critical for learning turbulence distributions are preserved even on coarser grid resolutions (Mortezazadeh et al., 2022). The solver uses an LES closure with the standard Smagorinsky Subgrid-Scale model (Smagorinsky, 1963), where the eddy viscosity is computed as $\nu_t = (C_s \Delta)^2 |\overline{S}|$, with the Smagorinsky constant C_s set between 0.1 and 0.24 depending on the flow regime and the filter width Δ derived from the local grid volume (Qin et al., 2025a). The CityFFD has been explored and validated against a series of studies ranging from experiment and simulations and has been integrated into Weather Research and Forecasting Model (WRF) as well as urban energy model CityBEM (Katal et al., 2019; Wang et al., 2023).

Building cluster geometries for all cities used are selected around the world and were downloaded from the CityFFD platform (https://cityffd.com). Later they are converted to STL format while maintaining 1 m resolution. These geometries encompass a broad spectrum of urban forms, ranging from low-rise residential neighborhoods to high-rise metropolitan clusters, thereby ensuring morphological diversity and heterogeneity. For each city, the simulation domain covers 6 km × 5 km × 2 km with a fine resolution of 4 m × 4 m × 1.5 m near the central building clusters, resulting in approximately 40 million grid cells per case as seen in Figure 1 (Qin et al., 2025b). The extended domain with large surrounding buffer zones ensures that the atmospheric boundary layer develops fully upstream and that building wakes dissipate without boundary interference. All converged simulation results were finally cropped to 1.2 km × 1.2 km × 240 m subdomains centered on the building clusters and resampled to 8 m × 8 m × 3 m grid spacing, resulting in tensors of shape (150, 150, 80). The way of cropping enables the capture of the turbulent flow within the urban canopy while excluding peripheral buffer zones that would waste model capacity on relatively uniform approach flow (Qin et al., 2025). The inlet wind blows from west to east at 4 m/s with a power-law profile:

$$u(z) = u_{ref}\left(z/z_{ref}\right)^{\alpha}$$

where $u_ref$ = 4 m/s at z_ref = 10 m and the roughness exponent α = 0.15 for suburban terrain. Building surfaces are set to 40 °C and the ambient air temperature to 30 °C. The simulation time step is $\Delta t = 0.5s$ with CFL ≈ 0.5. For each city, 500 instantaneous snapshots are saved at 5 s intervals, ensuring temporal decorrelation between successive samples. This sampling interval exceeds the characteristic time for large-scale coherent structures to pass a fixed point, so that each snapshot constitutes a statistically independent realization of the transient flow field.

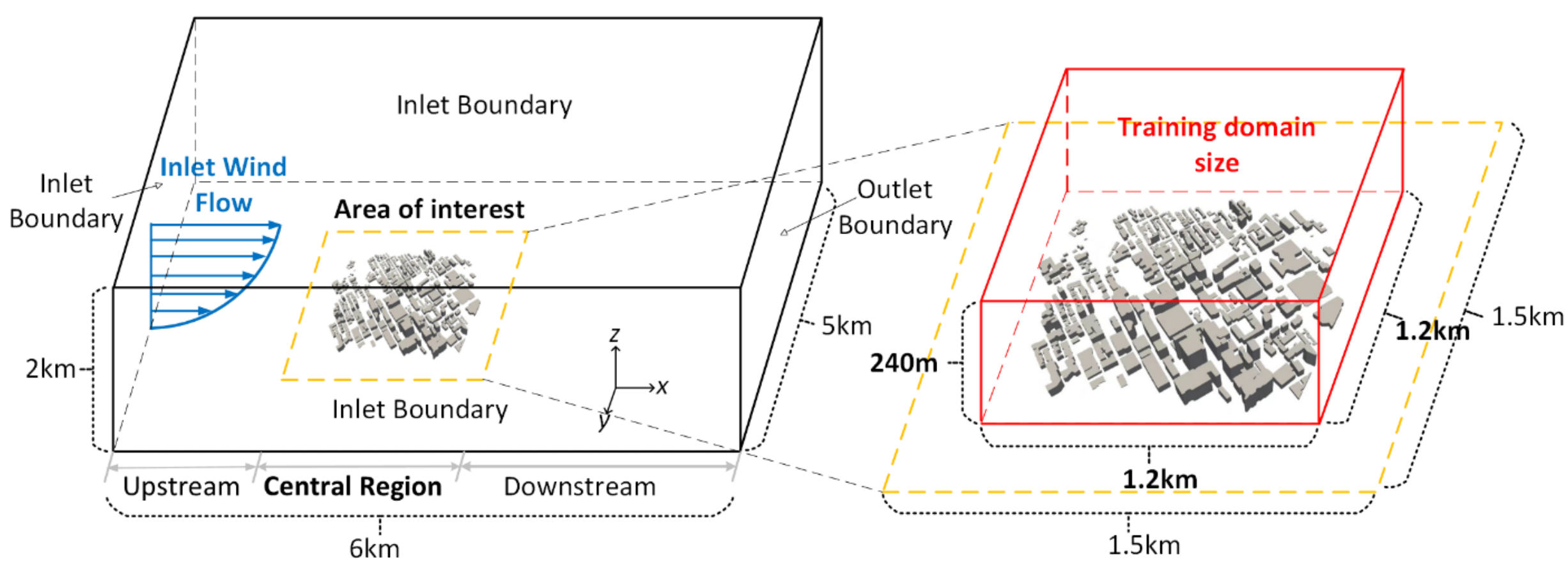

*Figure 1 Computational Domain for CityFFD simulation (left) and target area for model training (right)*

The central region of interest for CityFFD simulation (1.5 km × 1.5 km) includes buildings ranging from low-rise residential areas to skyscrapers up to more than 200 m in height, based on realistic urban morphologies that are selected. The dataset of the urban configurations is shown in Figure 2a. The diversity of urban morphologies is crucial for learning a generalized model, as different building layouts produce distinct flow regimes — channeling, downwash, and wake interference — that must all be represented during training to avoid overfitting to narrow patterns (Javanroodi et al., 2022). Due to this, the dataset is composed of 20 diverse and realistic urban morphologies for which 500 instantaneous flow field 3D snapshots were extracted.

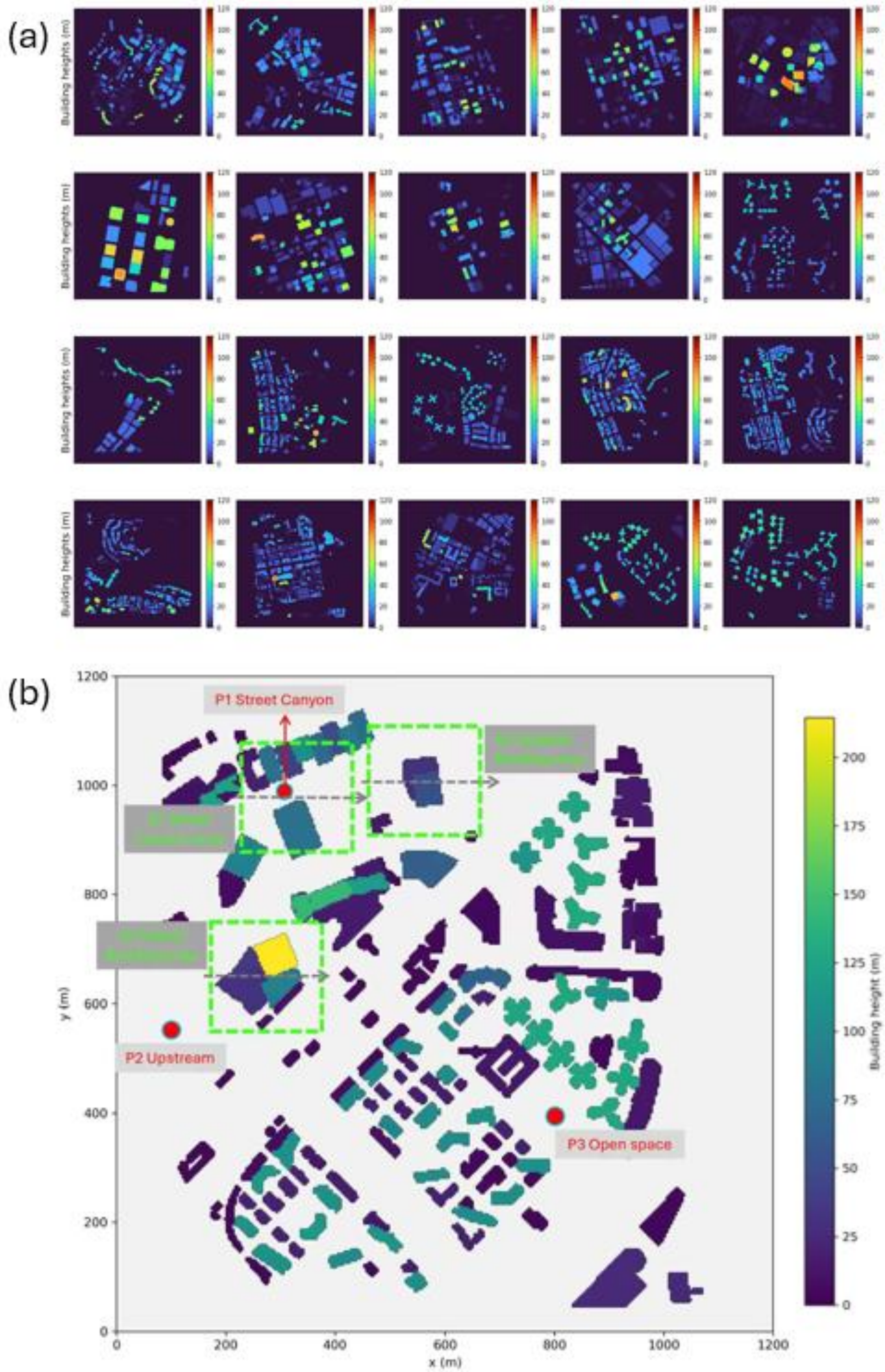

*Figure 2 (a) Urban morphology of all cases in the dataset colored with height (b) Urban morphology of the testing case and focused areas for validation*

For training and testing, dataset is partitioned into 17 training cities and 3 test cities. Rather than relying on a traditional validation set evaluated solely by loss magnitude, we implemented a robust progressive verification strategy, which is generating the snapshots based on checkpoints for each 2000 steps, to view the actual performance improvement from visual inspection as well as statistical censure, since most of the L2 losses used in training only reflects the general or averaged trend in probabilistic generative modelling, which can be insufficient since the it might hide the trade-off between sample quality and distribution coverage. For example, the accuracy or the quality of the generated results will keep progressing despite that the loss is already hitting the "plateau" (Sajjadi et al., 2018). In Figure 2b, the testing case study that is focused on this paper is presented. In order to facilitate the validation process of Section 3, and present results that reveal the accuracy of CFM in local scales, three diverse areas are chosen (see A1, A2, A3 in Figure 2b) and three district location for the vertical profiles (P1, P2, P3 in Figure 2b).

## 2.2 CFM framework

Flow Matching (FM) offers an alternative simulation-free generative framework for training Continuous Normalizing Flows (CNFs), which define a time-dependent vector field $v_t(x)\colon [0,1] \times \mathrm{D} \to \mathbb{R}^{d}$ that is calculated by Ordinary Differential Equation (ODE) solvers (Albergo & Vanden-Eijnden, 2022; Lipman et al., 2023). Such vector field governs the flow $\phi_t(x)$ of the probability path, which leads the samples from a source distribution $p_0$ (typically Gaussian noise) to a desired data distribution $p_1$ (Lipman et al., 2023, Tong et al., 2023). Mathematically, between two distributions, exists a "ground truth" vector field $u_t(x)$ that represents the mathematically perfect set of directions required to move noise into data along a specific probability path. The goal of FM is to approximate the parameterized vector field $v_\theta(t,x)$ that is trained by a Neural Network (NN) to the ground truth vector field $u_t(x)$ by L2 loss of $\mathcal{L}_{\mathcal{FM}}(\theta) = \mathbb{E}[\|v_\theta(t,x) - u_t(x)\|^2]$, where the expectation is taken over $t \sim U[0,1]$ and $x \sim p_t(x)$. In practice, directly optimizing this objective is often intractable because the marginal probability path $p_t$ and its corresponding vector field $u_t$ are unknown for complex distributions.

CFM introduces a conditioning variable $z$ (typically a sample pair $(x_0, x_1)$ from the source and target distributions) (Tong et al., 2023). By defining a conditional probability path $p_t$and a tractable conditional vector field $u_t(x|z)$, the CFM objective is formulated as also L2 loss between the parameterized vector field trained with a NN:

$$L_{CFM}(\theta) = E\left[\left||v_\theta(t,x) - u_t(x|z)|\right|^2\right]$$

Prior research has demonstrated that the gradient of the CFM objective is equivalent to the gradient of the original FM objective, provided the marginals are consistently defined. A widely adopted choice for the conditional path is the Optimal Transport (OT) displacement map, which utilizes a linear interpolation: $p_t(x|x_0, x_1) = N(x|(1-t)x_0 + tx_1, \sigma^2 I)$, The corresponding conditional vector field is simplified to: $u_t(x|x^0, x^1) = x^1 - x^0$ (Tong et al., 2024). This formulation ensures that the model learns to push samples along straight-line trajectories, which significantly improves sampling efficiency by allowing for larger steps in the ODE solver during inference. Unlike standard diffusion models that simulate a noising process through stochastic differential equations, CFM directly learns the velocity field of an ODE that transforms noise into data, yielding straight optimal transport paths that require fewer integration steps for generation (Lipman et al., 2023; Tong et al., 2024).

The architecture of the model is presented in Figure 3. The model inputs in total have 15 channels, which involves noisy states, conditions, and target states of instantaneous flow field snapshots. The original

Gaussian noise state $x_t$ that has the same shape as target states contributes 4 channels and serves as the starting point for inference which can be seen as the reverse of model training. The conditions comprise 7 channels: a building binary mask (1 ch), a Signed Distance Field (SDF) encoding the distance from each voxel to the nearest building surface (1 ch), the time-averaged steady-state flow field from CityFFD (4 ch: *u*, *v*, *w*, *T*) and a scalar roof height map (1 ch). An example of the conditional inputs is presented in Figure 4. The target states of transient snapshots have also 4 channels of wind velocities and temperature. The conditioning variables are static and non-trainable — their values remain fixed throughout training; only the network parameters are updated during backpropagation.

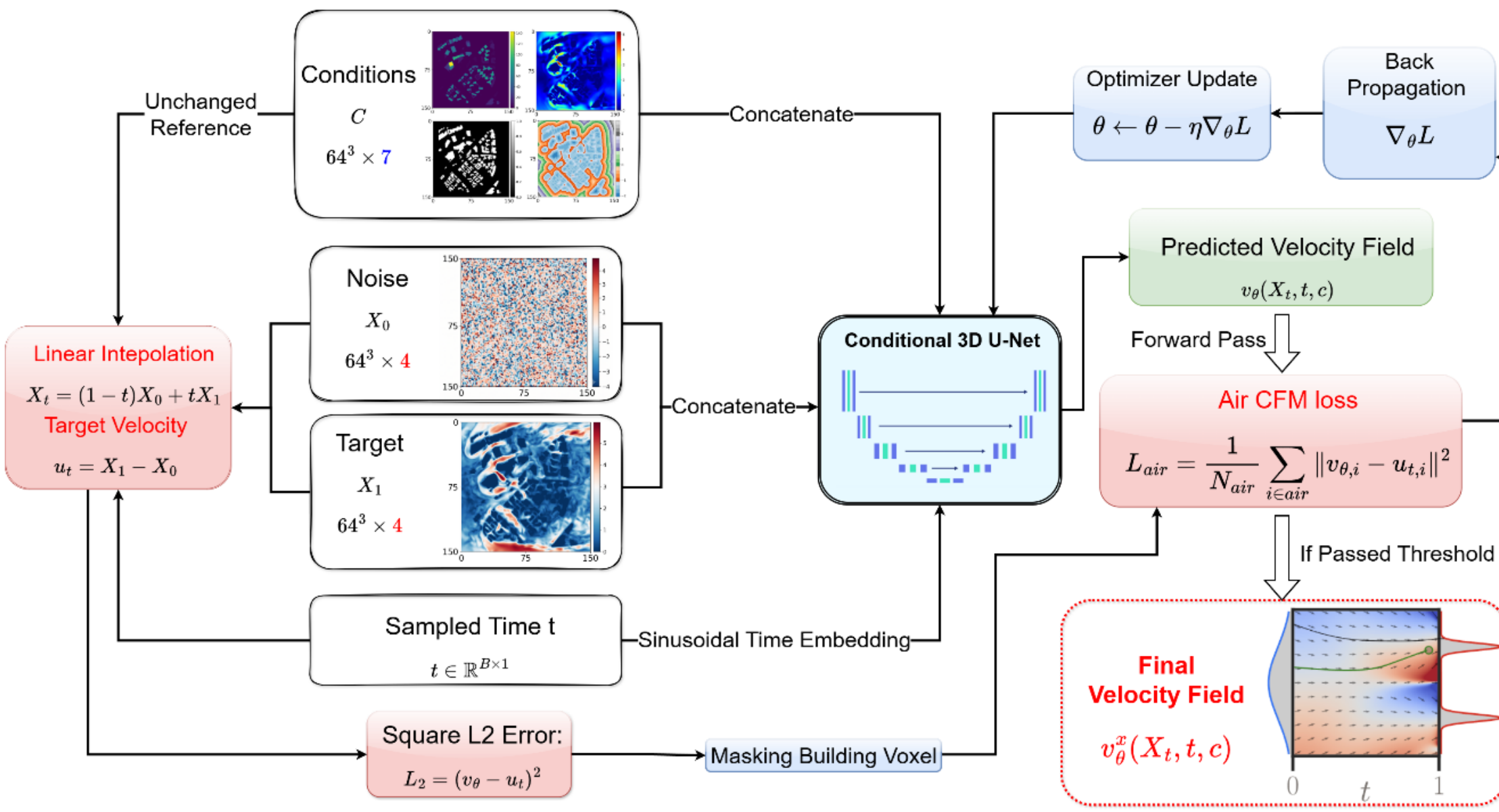

*Figure 3: CFM training and generation pipeline. Left: conditioning inputs concatenated with noisy state; center: U-Net predicts velocity field V_\theta; right: training target and generation.*

As shown in Figure 3, the backbone architecture is a 3D convolution U-Net with a base channel width of 64 and channel multipliers of (1, 2, 4, 4) for 4 levels of channels, resulting a bottom layer of 256 channels and $8^3$ of spatial resolution. Each hierarchical level includes two residual blocks for stable training as well as vector field continuity. Self-attention is added at the bottom layers with resolutions of 8 and 16 to ensure general global context modeling and long-range dependency capture in the flow field. The continuous timestep $t$ is projected into a sinusoidal positional embedding and processed by a Multi-Layer Perceptron with SiLU activations; this embedding is added to the feature maps in every residual block. The decoder uses skip connections from the encoder and nearest-neighbor up sampling followed by convolution. The final output is decoded and projected to 4 channels matching the velocity and temperature fields$u, w, v, T$.

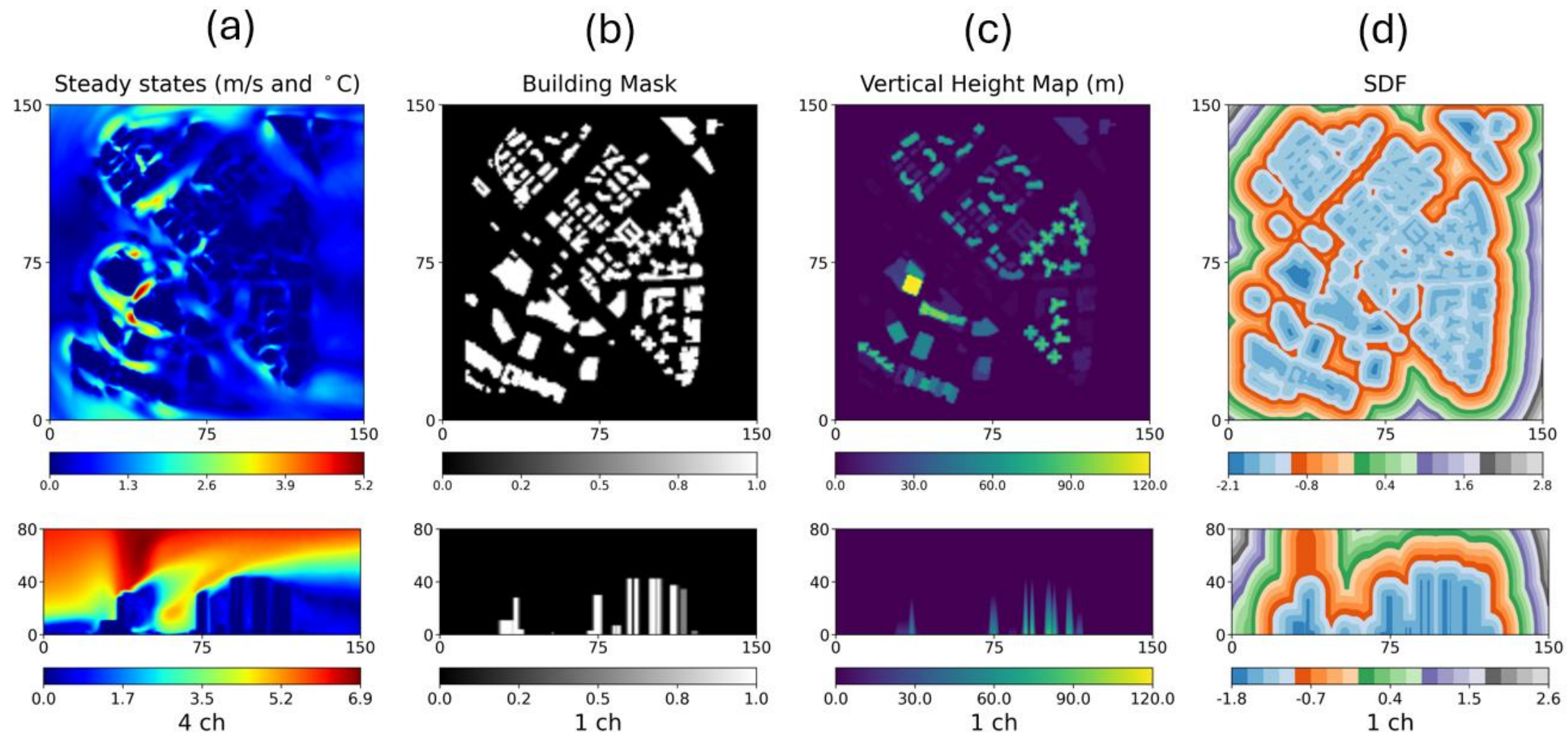

*Figure 4: All Conditional inputs: (a)Time-Averaged Mean Flow Field; (b) Building Binary Masks and (c)Building Roof Height in meters and (d) Signed Distance Function of the Buildings*

The training objective uses normal L2 loss as other CFM models, but masks are adopted to separate the computational domain into air and solid regions based on building occupancy for better forcing the model on the airflow, not the building grids. The air mask $M_{air}$ zeroes out velocity predictions within building volumes and excludes these regions from gradient computation. The masked flow-matching loss is formulated as

$$L_{CFM} = E_{t,x_0,x_1}[\Sigma M_{air}(x)||v_\theta(t,x_t,c) - u_t(x_t|x_0,x_1)||^2]$$

, normalized by the number of air voxels $N_{air}$ rather than total domain size to maintain consistent loss magnitudes across cases with varying building density. An additional weak regularization term penalizes non-zero predictions in solid regions. Training uses the Adam optimizer with a peak learning rate of $2\times10^{-4}$ and linear warmup over 5000 steps, for a total of 400,000 iterations. Exponential Moving Average of model weights with a decay of 0.9999 is maintained and used for inference. To verify the training status, qualitative results were generated every 2,000 steps across a total of 400,000 iterations. This approach is widely recommended for probabilistic generative models, as it ensures that the model captures the desired data distribution beyond simple loss convergence like L1 or L2 loss target, but also achieving tangible generative capability improvement both visually and statistically (Goodfellow et al., 2014; Johnson et al., 2016).

All flow field channels are normalized using z-score standardization computed across all training, validation, and testing cities. This transformation aligns the target distribution with the Gaussian noise prior $p_0$ ~ N (0, I), minimizing the optimal transport distance and allowing the network to prioritize learning intricate flow topologies rather than global magnitude shifts (Tong et al., 2023). Geometrical inputs (height, SDF, coordinates) are also scaled to [0, 1] or [−1, 1] through relative normalization such as MinMax and Standard scaling. In the U-net architecture, each hierarchical level contributes to the representation of flow features at different spatial scales. The first level, with 64 channels, primarily captures the fundamental geometric boundaries of the computational grid, including building edges and local solid-fluid interfaces. The second level, with 128 channels, further extracts localized shear-flow features, particularly those associated with street-canyon effects and near-building velocity gradients. At the third level, the 256-channel representation enables the model to encode intermediate-scale vortical structures that emerge from flow separation and recirculation around buildings. The deeper fourth and fifth levels, each with 512 channels, aggregate broader flow information, including Reynolds-stress-related patterns and large-scale advective structures. These deeper representations form the core of the model's first dominant mode, allowing the network to integrate local geometric constraints with global urban-flow organization.

## 2.3 Patch-based 3D generation and domain reconstruction

High-resolution 3D generative modeling has inherent scalability limitations, which are widely reported to hit GPU memory limits, especially when operating directly on full volumetric tensors rather than compressed representations in latent space (Miller et al., 2023; Uzunova et al., 2019).This study adopts an straightforward, practical, yet effective strategy of localized patches to address the hurdle without the needs of a pretrained encoder-decoder neural network to convert the full domain into latent space, which entails the possibility of blurred local structures when decoding (Bredell et al., 2023; Rombach et al., 2022). Apart from optimizing the model configurations to lower the parameters during training, the localization strategy was adopted on our single NVIDIA RTX 6000 pro ADA 48 GB graphic card. We crop the full tensor shaped as (150,80,150) to $64^3$ voxel patches, which divides the whole domain into $3 \times 3 \times 2 = 18$ 3-dimensional patches with overlaps and data format of FP32. During training, the overall GPU ram usage for batch size of 4 is around 44100 mb, which is around 44 Gb, taking full advantages of our RTX 6000 pro ADA.

While this patch-based approach enables tractable training, it introduces a fundamental challenge during inference: independently generated patches are not spatially coherent and naively assembling them would produce visible discontinuities at patch boundaries. To resolve this, we employ a shared noise field initialization strategy combined with per-step weighted linear merging scheme during ODE integration to stitch the generated patches back to full domain. This process is displayed in Figure 4. These patches, rather than generated from independent noise samples that have different variations, we initialize a global Gaussian noise field $\mathcal{Z}: x_0 \sim N(0, I)$, whose dimensions matching the complete target domain (C, D, H, W), which in this case is (4, 150, 80, 150). This shared initialization ensures that overlapping regions of different patches begin from identical noise realizations, providing a mathematical foundation for coherent reconstruction from the very beginning.

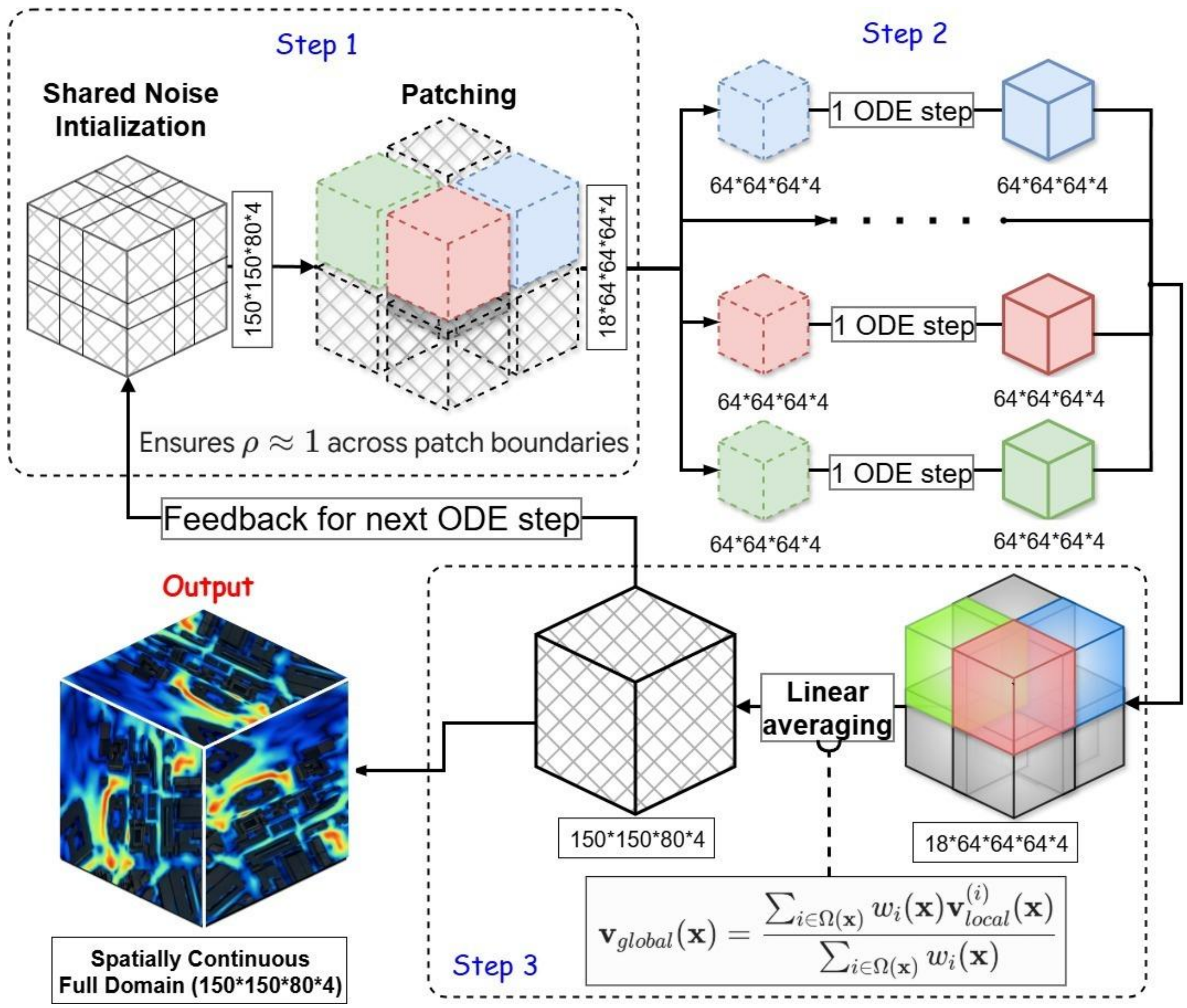

*Figure 4 Schematic of the patch-based domain reconstruction: (a) overlapping $64^3$ patches tiling the full domain; (b) shared-noise initialization; (c) per-step linear weighted averaging producing the full-domain field.*

The inference process is framed as an iterative synchronization loop within the Ordinary Differential Equation (ODE) solvers as shown in Figure 4. At each repetitive ODE timestep $t_k$ , there are 3 steps that are done: First, the full domain is decomposed into 18 overlapping patches with horizontal stride $s_hon$ = 43 and vertical stride $s_ver$ = 16, yielding coordinates of the patches to be $\mathrm{P} = \{(\mathrm{d}, \mathrm{h}, \mathrm{w}) | \mathrm{d} \in [0, \mathrm{D} - \mathrm{p}]_s, \mathrm{h} \in [0, \mathrm{H} - \mathrm{p}]_s, \mathrm{w} \in [0, \mathrm{W} - \mathrm{p}]_s\}$, where $s < p$ to ensure the overlap; Second, each patch will undergo a single-time transportation following the trained intermediate vector field $v_\theta$, which is calculated by fourth-order Runge–Kutta (RK4) ODE solvers; The third step requires remerging with weighted linear averaging over the overlaps while the weights of each pixel are based on their distance between the patch core, where a raised 3D cosine kernel is applied as the weight mask to ensure smooth spatial transitions; After all 3 steps are passed, the merged global field is advanced to the next state $t_{k+1}$. Such loops will be implemented multiple times to finally get the correct and spatially smoothed instantaneous flow field snapshots at $t_n$. This per-step integration ensures that spatial coherence is maintained throughout the entire denoising trajectory, rather than being applied as a post-processing step. In practice, fewer than 10 ODE steps are sufficient to produce physically plausible, LES-quality transient flow states (Lipman et al., 2023; Lu et al., 2022).

As regards remerging of the generated patches, while some paper uses variance preserving averaging (the square-root method) that is mathematically "safer" for independent signals since it mathematically retains the energy level across patches, it is very sensitive to phase shifts or tiny numerical disagreements. If patches disagree by even a few pixels on where a "swirl" is located, this method amplifies that disagreement, leading to the "shattered" look or the hard lines. In standard signal processing, linear averaging is often avoided due to the variance collapse theorem. For $N$ independent predictions $\mathrm{v}^{(\mathrm{i})}$ , the variance of their average decreases by a factor of $N$, which would normally lead to 'artificial smoothing', which can be stated as:

$$\mathrm{Var}\left(\frac{1}{N}\sum_{i}^{N} v^{(i)}\right) = \frac{1}{N^2}\sum Var(v^{(i)}) = \frac{N\sigma^2}{N^2} = \frac{\sigma^2}{N} \#(1)$$

However, since the shared noise initialization is adopted, the patches aren't independent—they are structurally aligned from the very first step with a a correlation coefficient $\rho \approx 1$ across patch boundaries, which bypassed the "variance collapse" assuming neighboring patches $v_i$ and $v_j$ are uncorrelated. Because the overlapping regions of Patch A and Patch B share the same latent starting point, they are exposed to identical initial noise as well as the same global SDF and mean-flow conditions. Since the neural network operates deterministically, identical input noise should produce highly consistent outputs. As a result, the predicted velocity fields in the overlapping area are expected to be nearly the same, such that $v^{(1)} \approx v^{(2)}$, if correlation factor $\rho = 1$. Then the linear averaging acts as a "soft consensus." It trusts that the shared noise has already done the work of aligning the structures, so it simply smooths out the tiny numerical residuals at the boundaries:

$$V_{avg} = \sum \omega_i \times v^{(i)}, \text{ where } \sum \omega_i = 1$$

## 2.4 Evaluation metrics

The evaluation framework is designed to validate the generative model across a robust system of statistical stringency and errors for some localized metric, from the convergence of the generated transient states to local flow profile accuracy at characteristic urban locations. For most building engineering and urban planning tasks, statistical wind properties — means, variances, PDFs — are more representative and practically useful than full temporal trajectories, with detailed time series reserved for wind-induced structural vibration and short-term pollutant dispersion analyses (Tominaga et al., 2008; Cao et al., 2024; Zhang et al., 2015).

a. Convergence of generated ensemble

Firstly, before other metrics, the statistical convergence of the generated ensemble is calculated, which determines the minimum ensemble size required to stabilize first and second order statistics. The convergence of the running mean μ and standard deviation σ is monitored as a function of ensemble size n, which can be calculated as:

$$E_\mu(n) = \left| \frac{1}{n} \sum_{i=1}^{n} \phi_i\,(\mathbf{x}) - \bar{\phi}_{LES}(\mathbf{x}) \right|$$

$$E_\sigma(n) = \left| \sqrt{\frac{1}{n-1} \sum_{i=1}^{n} (\phi_i(\mathbf{x}) - \mu_i)^2} - \sigma_{LES}(\mathbf{x}) \right|$$

And the threshold of convergence is set to be 0.01% residuals for the rate of change in $\mu$ and $\sigma$ for each variable.

b. Flow field statistics

Secondly, to validate whether the model can reproduce the same first order means velocity and temperature map against same number of randomly selected CityFFD snapshots, we adopted the follow equations to calculate the mean, where $N_g$ and $N_r$ will be the same value due to ensemble size:

$$\bar{\mathbf{u}}_{gen}(\mathbf{x}) = \frac{1}{N_g} \sum_{i=1}^{N_g} \mathbf{u}_i^{gen}\,(\mathbf{x})$$

$$\bar{\mathbf{u}}_{LES}(\mathbf{x}) = \frac{1}{N_r} \sum_{i=1}^{N_l} \mathbf{u}_i^{LES}\,(\mathbf{x})$$

Thirdly, the second-order statistics are implemented to validate the model's ability to capture turbulent fluctuations. Standard deviation between the generated snapshots for testing cities are obtained to quantify the amount of fluctuations to understand some of the regional strength or weaknesses of the generated results. The standard deviation for a variable u is defined as:

$$\sigma_{\mathbf{u}}(\mathbf{x}) = \sqrt{\frac{1}{N} \sum_{n=1}^{N} \left\| \mathbf{u}^{(n)}(\mathbf{x}) - \bar{\mathbf{u}}(\mathbf{x}) \right\|^2}$$

And also Turbulent Kinetic Energy (TKE), which translates those fluctuations directly into physical fluid dynamics to assess whether the generative model respects the fundamental laws of conservation of energy and fluid motion, and it is defined as:

$$TKE = \frac{1}{2}\left(\overline{(u')^2} + \overline{(v')^2} + \overline{(w')^2}\right), \text{ where } u' = u - \bar{u}.$$

Fourthly, the Probability Density Functions curves are constructed by aggregating velocity and temperature samples across all spatial grid points and all snapshots, providing a rigorous test of distributional accuracy including tail behavior and multimodality. To estimate the continuous probability density function $\hat{f}(x)$ for each velocity component and temperature dataset, we employ Kernel Density Estimation (KDE). This non-parametric approach smooths the discrete empirical sample distribution by centering a standard Gaussian kernel function over each of the randomly sampled data points $x_i$:

$$\hat{f}(x) = \frac{1}{nh}\sum_{i=1}^{n}\frac{1}{\sqrt{2\pi}}\exp\left(-\frac{(x-x_i)^2}{2h^2}\right)$$

where $x$ represents the evaluation grid points, and $h$ denotes the smoothing parameter or bandwidth. To ensure an optimal balance between over-smoothing (bias) and under-smoothing (variance), the bandwidth $h$ is dynamically selected according to Scott's Rule:

$$h = \sigma \cdot n^{-1/5}$$

Here, $\sigma$ represents the sample standard deviation of the empirical data, and n is the total number of sampled data points utilized in the kernel evaluation.

Finally, the local wind profile is performed at three representative probe locations within the test domain: an upstream inflow site (P1) to validate the preservation of the boundary layer profile, a street canyon site (P2) to assess recirculation and shear, and an open area (P3) to evaluate ventilation in sheltered open spaces. At each location, vertical profiles of ensemble mean, and standard deviation are compared against CityFFD references across all 80 height levels. First, for a given location, N number of values will be recorded for one height, which are used to calculate the mean and standard deviation for this height:

$$\mu = \frac{1}{N}\sum_{i=1}^{N} P_i\,, range \;=\; [Min(P_i), Max(P_i)]$$

Where N represents the number of snapshots in the ensemble, $P_i$ stands for the variable values $u, w, v, T$ for a snapshot at given x and y coordinate and height. At each designated location, the script samples a column of data upwards through the entire vertical axis from height $z = 0\ m$ to $z = 240\ m$.

c. Error metrics:

Several metrics are used to provide statistical errors regarding to the quality of the generated ensemble. Root Mean Square Error (RMSE)and Mean Absolute Error (MAE), along with their range-normalized value which is divided by their own mean values are used to measure the average error, where lower magnitudes reflect good fidelity to the CityFFD reference. These metrics are defined as follows for a variable u:

$$\text{RMSE} = \sqrt{\frac{1}{N}\sum_{i=1}^{N}\left(\mathbf{u}_{gen,i} - \mathbf{u}_{LES,i}\right)^2}$$

$$\text{MAE} = \frac{1}{N}\sum_{i=1}^{N}\left|\mathbf{u}_{gen,i} - \mathbf{u}_{LES,i}\right|$$

$$\text{NRMSE} = \frac{\text{RMSE}}{u_{truth}^{max} - u_{truth}^{min}} \times 100\%$$

$$\text{NMAE} = \frac{\text{MAE}}{u_{truth}^{max} - u_{truth}^{min}} \times 100\%$$

Where $N$ is the size of the size of the ensemble, which is the number of snapshots generated. The same number of CityFFD reference snapshots are selected randomly from the testing dataset. And $u$ can be the value for each channel, namely wind velocity and each direction, $u$, $w$, $v$, as well as temperature T.

For localized analysis that aims in the 200-meter area in urban space, Averaged Bias and Pearson Correlation coefficient is used to measure the similarity between the generated values and CityFFD reference. The mean Bias is defined as:

$$\text{Bias} = \frac{1}{n}\sum_{i=1}^{n}(y_i - x_i)$$

Where positive magnitude suggests an overestimation ad negative magnitude an underestimation of the overall wind speed or temperature.

Since Pearson correlation coefficient is a linear statistical metric used to measure the strength and direction of the relationship between two continuous variables, it is defined as:

$$r = \frac{\sum_{i=1}^{n}(x_i - \mu_x)\left(y_i - \mu_y\right)}{\sqrt{\sum_{i=1}^{n}(x_i - \mu_x)^2}\sqrt{\sum_{i=1}^{n}\left(y_i - \mu_y\right)^2}}$$

While a correlation of r = 0 means $X$ provides zero predictability about Y, a perfect negative correlation of $r$ = -1.0 means $Y$ decreases in exact proportion as X increases, aligning the data points perfectly on a downward-sloping straight line.

# 3. Results

This section presents the comparable results of the generated ensemble against CityFFD LES simulation and discussion based on the comparison. We first evaluate the number of snapshots that needs to be generated to be statistically converged. Then we exhibit samples of generated snapshots as direct visualization with zoomed in perspectives along the roof-height area and wake regions. Next, based on the converged ensemble, we evaluate the quality of the generation in terms of first-order flow statistics such as mean flow pattern, second-order flow statistics such as standard deviation and TKE, global Probability Density Function (PDF) for all 4 variables, as well as local wind profiles for 3 characteristic selected spots in the test city to view the variable gradients for vertical span.

## 3.1 Convergence of generated ensemble

Our generative surrogate is used to produce ensembles of instantaneous flow field snapshot, which is inherently chaotic and independent from each. Thus, calculating the convergence of the ensemble can verify that enough snapshots are collected to reach a steady, repeatable state and the calculating method can be referred in 2.4.

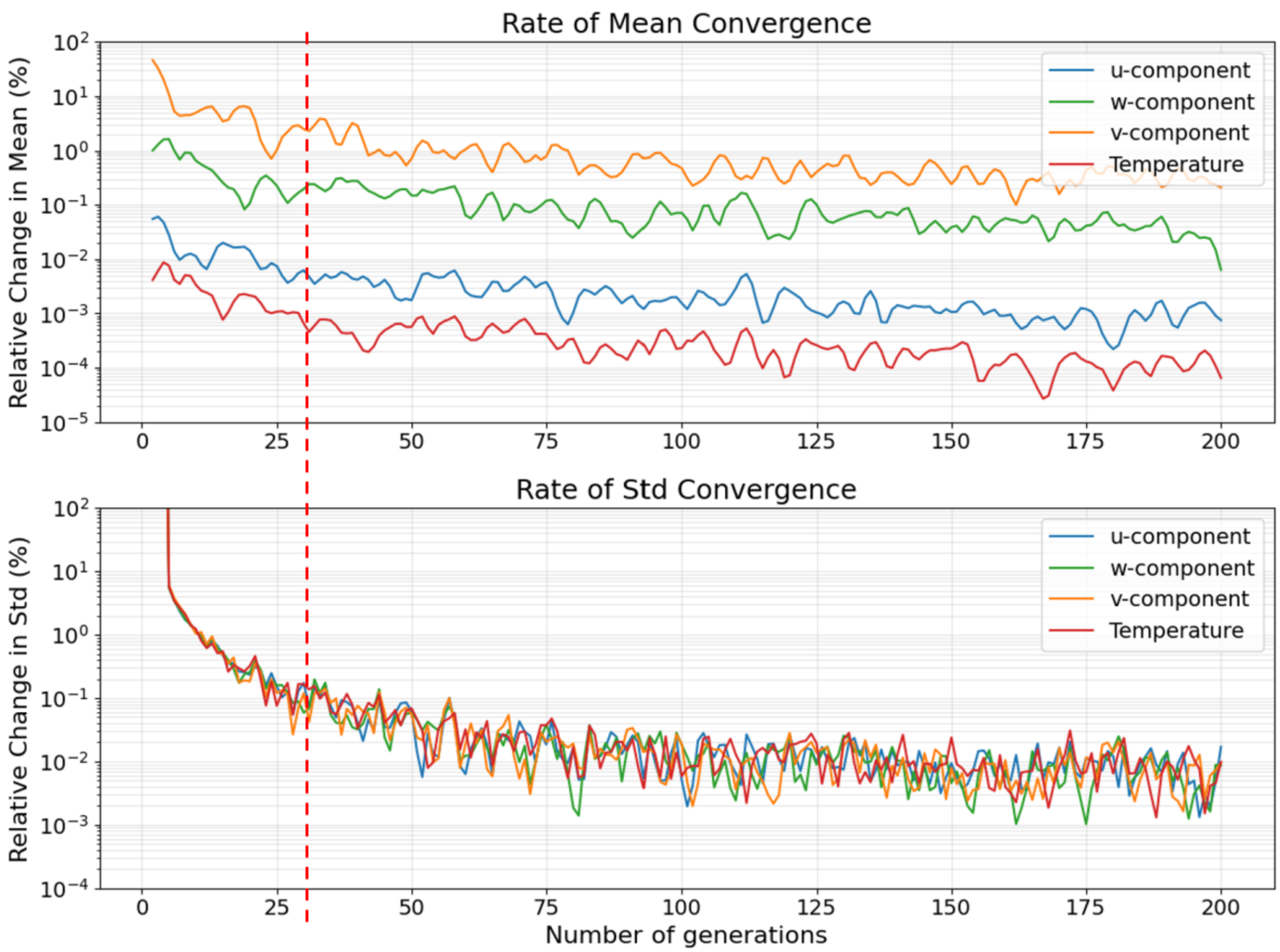

*Figure 5: Convergence of mean and standard deviation for the ensemble of 200 generated snapshots*

As it is shown in Figure 5, The rate of change in the running mean stabilizes around N = 35, with a change rate of less than 0.01% hereafter. However, the standard deviation change rate stabilize around N = 70 for it to be less than 0.01%. To cater both the numerical accuracy and inference speed for generating the whole ensemble, we choose the size of 35 snapshots as the research target. According to previous study, our choice is consistent with the broader turbulence literature because different statistics converge at different rates, with means stabilizing earlier than variances or standard deviations, and because practical ensemble design often accepts a compromise between strict statistical convergence and computational cost (Tempest et al., 2023)

## 3.2 Instantaneous flow generations

Because the proposed model is generative, the objective is not to replicate the reference snapshot on a pixel-by-pixel basis, but rather to synthesize independent, physically plausible realizations that share the same underlying statistical and structural properties as the ground truth. During inference, generating an instantaneous 3D urban microclimate snapshot takes around 10 to 13 seconds, which makes the total time for generating a statistically converged ensemble, namely 35 instantaneous snapshots, in less than 8 minutes. To better present the quality of the generated flow field, Figure 6a presents a visual comparison of 3 randomly selected generated snapshots against one also randomly selected CityFFD reference snapshots for the testing urban configuration in Figure 2b at vertical height of 3 m, while Figure 6b presents the vertical slice in the middle where x = 600 m. Despite the crop of patches processing through overlapping $64^3$ patches, the resulting fields are spatially continuous without any visible gaps or transition, demonstrating the effectiveness of the share-noise initialization and weighted linear merging technique during ODE integration steps. The model produces physically structured instantaneous fields that exhibit realistic turbulent features like building wakes, street canyon vortices and various sizes of recirculation zones without visible discontinuities at patch boundaries.

Crucially, the CFM samples successfully capture the macro-scale physical patterns dictated by urban geometry. In the wind magnitude fields (both horizontal and cross-sectional), high-velocity channels ($\sim 5.5 m/s$) are correctly positioned along wide avenues and open peripheral boundaries, while stagnation zones and low wind speeds $0$ to $2\ m/s$ are consistently generated within dense building clusters. Similarly, the temperature fields across all samples accurately reproduce the urban heat island effect, trapping higher temperatures ($36$ to $40°\mathrm{C}$) inside the street canyons while maintaining cooler ambient boundaries. The micro-scale differences observed among the CFM samples—such as the unique shapes of local turbulent eddies, varying detachment points of building wakes, and slight fluctuations in the thermal plumes—reflect the inherently chaotic and stochastic nature of turbulent fluid dynamics. Rather than a limitation, these variations demonstrate that the model has learned the true underlying distribution of the flow field, allowing it to generate diverse yet physically structured instantaneous fields that exhibit realistic features like street canyon vortices and various sizes of recirculation zones without visible discontinuities at patch boundaries.

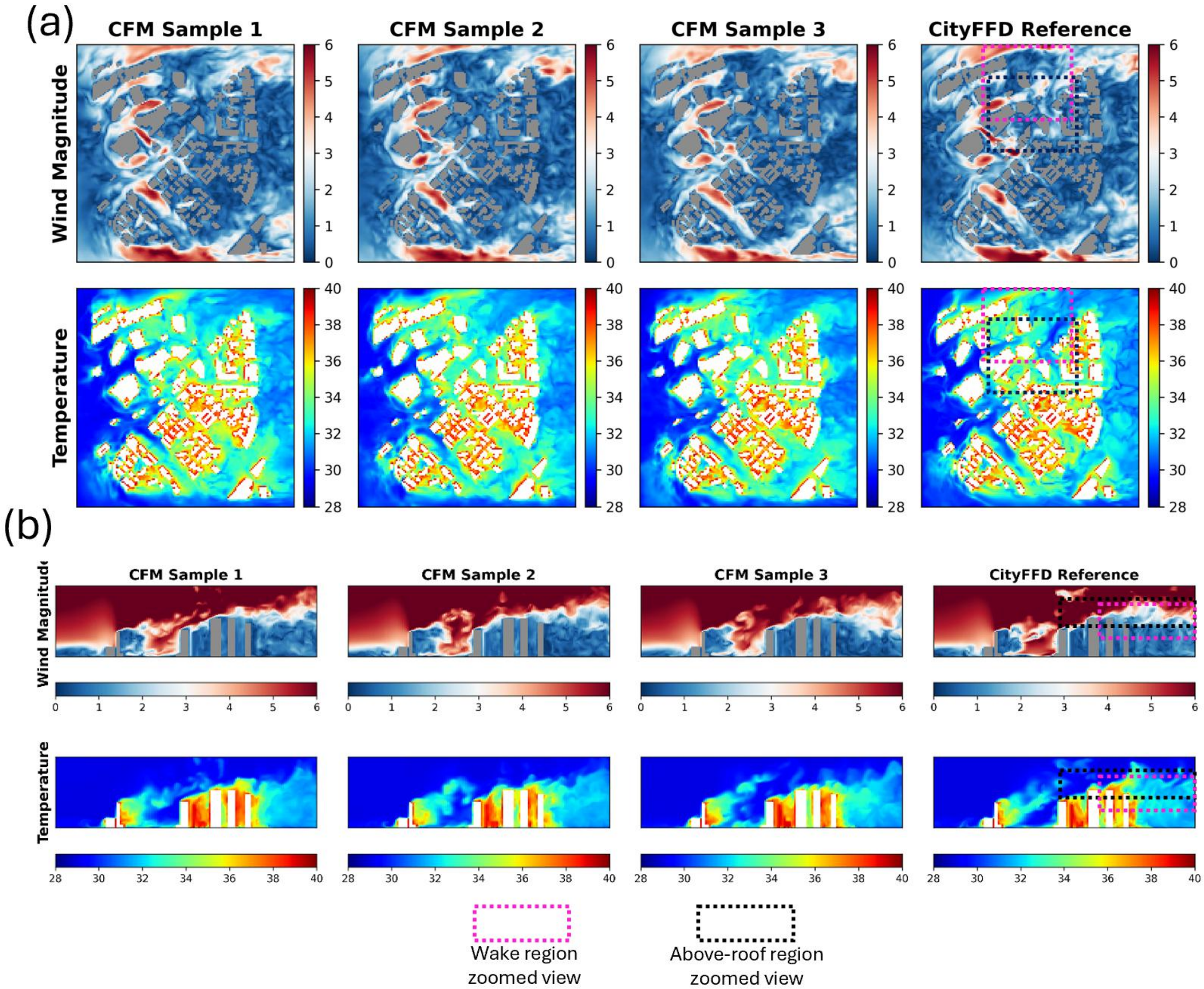

*Figure 6: Visual comparison of generated snapshots samples (first 3 columns) vs. random CityFFD reference snapshot (right column) at (a) horizontal z = 3 m and (b) xz vertical plane in the center*

To further quantify the differences of the generated from CFM Figure 7 presents a zoomed area above the roof region highlighted with the black box in Figure 6, and Figure 8 presents a zoomed region at the wake of the building highlighted with pink color in Figure 6. Regarding Figure 7, for the above-roof region, the vertical view shows how the flow accelerates over the roof edge and how the thermal plume rises above the buildings. The target has a sharper high-speed cap near the roof and a more distinct warm layer near the building edge. In the zoomed horizontal view at height of 100 m, the target also shows stronger spatial organization: high-speed channels and warmer patches are more coherent, while the samples generated are more diffuse. Sample 2 has the smallest horizontal wind bias, -0.026 m/s, but Sample 3 is visually close in the vertical wind profile. Temperature above the roof is less consistent: vertically, all samples are warmer than the target, while horizontally the mean bias is slightly cool, meaning the error depends strongly on local plume placement.

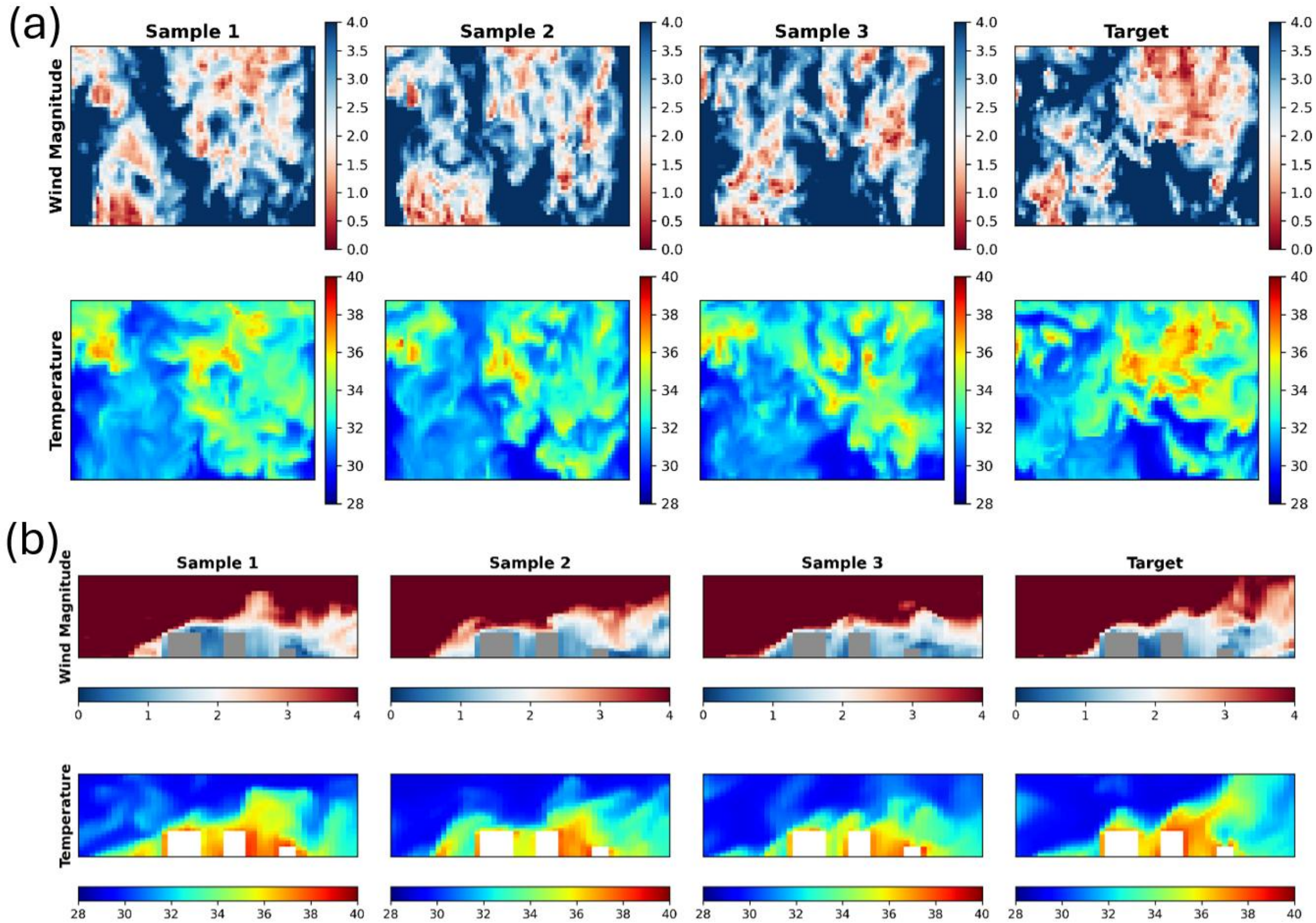

*Figure 7: Zoomed-in Visual comparison of above-roof region: generated snapshots samples (first 3 columns) vs. random CityFFD reference snapshot (right column) at (a) horizontal slice of z = 100 m and (b) xz vertical plane in the center*

And for the wake region, presented in Figure 8, the vertical view shows the downstream recirculation zone behind the building group. The target has a more structured wake, with alternating low-speed and high-shear areas. Samples 1 and 2 capture the broad low-speed wake but smooth out the sharper structures; Sample 3 tends to over-strengthen parts of the wake vertically. In the horizontal wake slice at z= 41, all samples underpredict wind speed relative to the target, with wind biases around -0.12 to -0.25 m/s, and the generated wakes look broader and less sharply bounded. Temperature in the wake is consistently warmer in the generated samples, especially near building edges, suggesting that the generated fields retain or spread heat more strongly than the reference. Overall, for the zoomed views, it is shown that the generated samples reproduce the amazing large-scale roof acceleration and wake deficit, but the reference has sharper local structures: stronger roof-edge gradients, more organized wake shear, and cooler downstream thermal pockets.

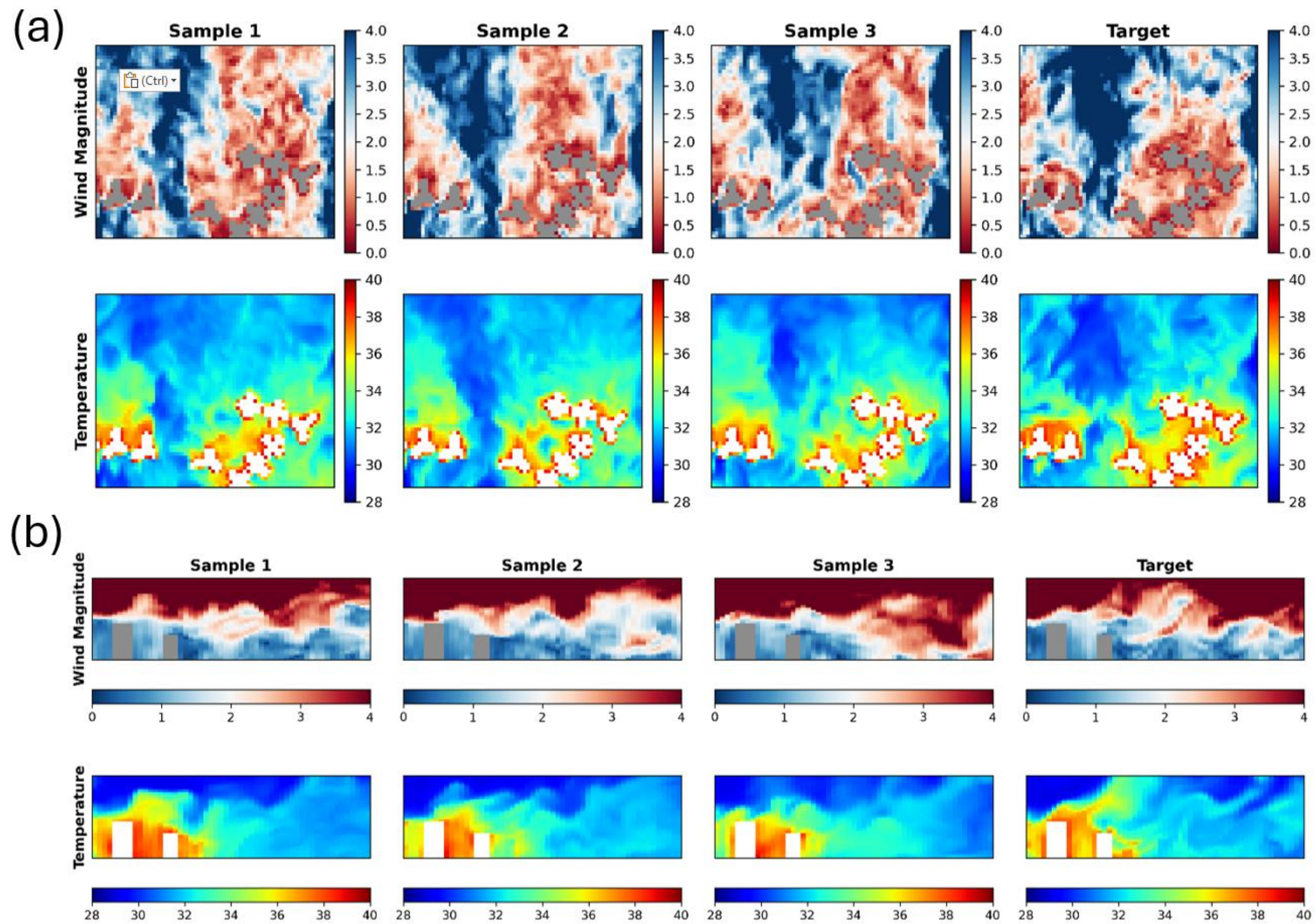

*Figure 8: Zoomed-in Visual comparison of wake region: generated snapshots samples (first 3 columns) vs. random CityFFD reference snapshot (right column) at (a) horizontal slice of z = 135 m and (b) xz vertical plane in the center*

## 3.3 Mean microclimate generation

The ensemble mean of the 35 generated snapshots is compared against the CityFFD time-averaged mean flow field to verify that the generative model reproduces the conditioning mean flow with sufficient accuracy. The spatial distributions of the mean velocity components (u, v, w) and temperature (T) across both horizontal (z = 3 m) and vertical (central yz-plane) slices demonstrate a high degree of qualitative and quantitative agreement between the generative output and the CityFFD LES reference (Figure 9).

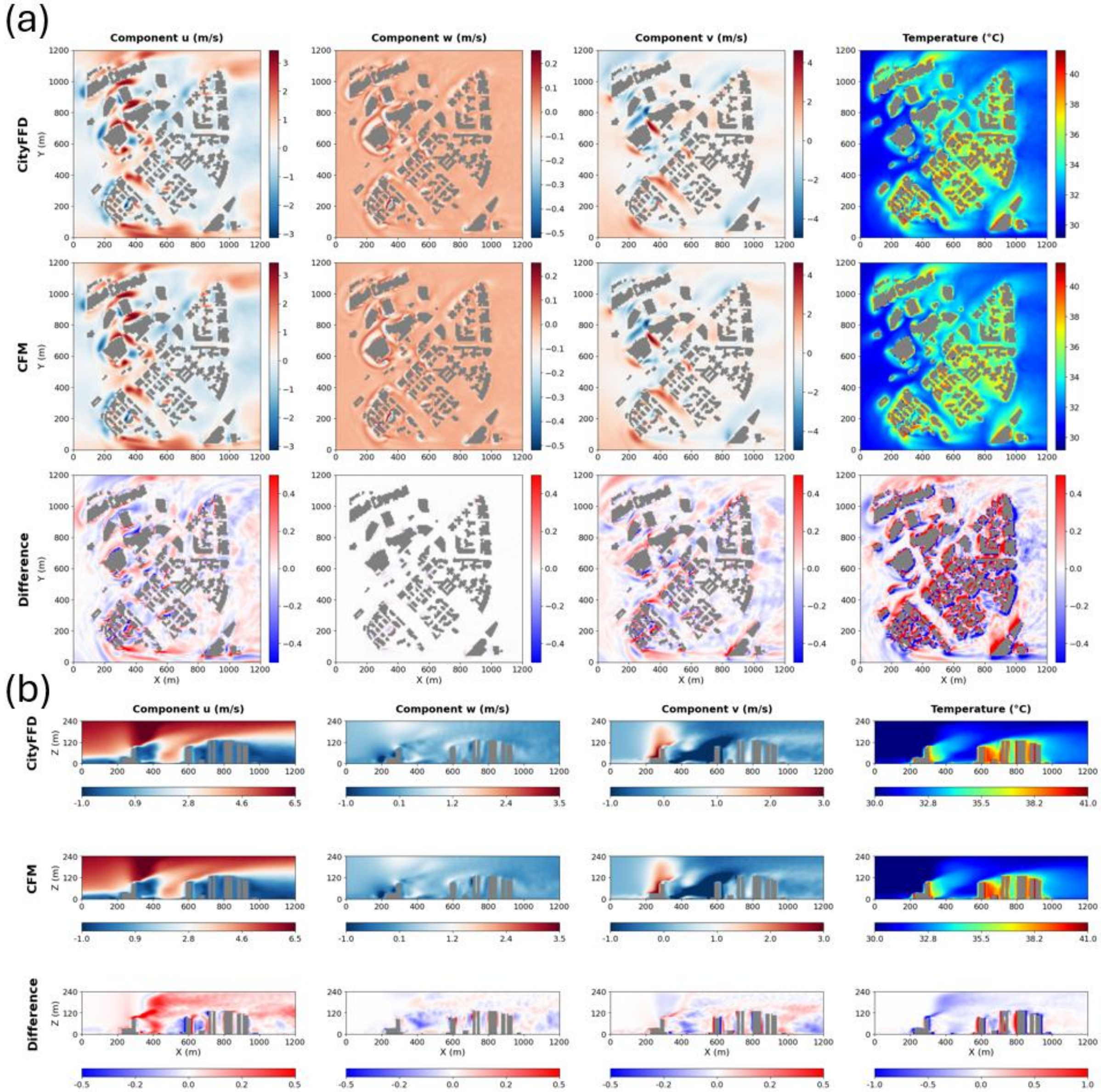

*Figure 9 Horizontal and vertical slices comparing the ensemble mean from 35 generated snapshots against CityFFD time-averaged mean flow. Top row: generated mean; middle row: CityFFD reference; bottom row: difference for (a) z= 3 and (b) central xz plane*

As seen in Figure 9, the mean flow field calculated by generated snapshots demonstrates exceptional fidelity in reconstructing both the building aerodynamic and thermal plume structures in a complex urban boundary layer. This strong macroscale agreement across both the velocity vectors $u, v, w$ and the thermal plume dispersion $T$ proves that the generative framework successfully learns the governing statistical mechanics of the fluid domain without suffering from excessive smoothing or structural distortion. The stagnation zones windward of the buildings and the high-velocity channeling effect in the narrow passages are preserved with minimal smoothing. The difference maps indicate that the largest discrepancies occur at the building interfaces, which attributes to the high gradients at the solid-fluid boundaries.

Upon looking at the details, the horizontal slices in Figure 9a reveal that the CFM model successfully reconstructs the intricate, geometry-dependent flow patterns nested within individual street canyons. Highly localized aerodynamic phenomena, such as the stagnation zones on the windward faces of the buildings and the high-velocity channeling effects through narrow urban passages, are preserved with remarkable precision. This agreement between the reference and generated results extends seamlessly to the scalar temperature fields, where the model cleanly maps the intense thermal trapping (38-41°C) inside dense building clusters alongside its subsequent downstream advection. By precisely capturing these sharp, localized spatial transitions in both the horizontal X-Y and vertical X-Z planes, the model demonstrates a robust capacity to handle complex bluff-body obstructions and secondary flow structures.

The vertical slices in Figure 9b, reveal that the CFM model accurately captures the separation of the boundary layer as it interacts with the high-rise structures. The wake regions and the recirculating vortices behind the buildings, critical locations for pollutant dispersion and pedestrian comfort, closely following the LES reference flow field. In the subplots of component u, the sharp vertical gradient separating the retarded canopy flow (visible in the blue-to-white range of -1.0 to 0.9 m/s) from the high-speed free stream (approaching deep red values of 6.5 m/s) is clearly observed. The corresponding Difference subplot for component u confirms that minor structural discrepancies, with maximum error around 0.5 m/s, are heavily localized exactly at the roof-level separation points and within the immediate leeward wakes of the tallest obstructions. Furthermore, the component w subplot demonstrates exceptional vertical momentum conservation; the model precisely maps pronounced windward updrafts reaching up to 3.5 m/s and subsequent leeward downdrafts dropping to -1.0 m/s, and the clean error map for w confirms this structural accuracy. Correspondingly, the temperature (°C) subplot highlights the successful reconstruction with minor error for high-speed wake region from the upstream tall building to the left around -0.5 °C. The model can capture localized heating that peaks near 41.0 °C and minimum temperature of 30.0°C from cooler ambient boundary layer. The difference subplot for temperature provides compelling visual evidence that geometric mismatches (constrained within a ±1.0 °C band) are strictly confined near the building surfaces such as walls and roofs, while the elevated plume in bulk air movement are reproduced correctly for open space between the buildings.

*Table 1 Error metrics for the ensemble mean flow field (35 generated snapshots vs. CityFFD time-averaged input as reference, where the reference wind velocity for u:4.3 m/s, w:0.23 m/s, v:0.3 m/s, and T: 31.95 ℃ ).*

| Variable | RMSE | MAE | NRMSE | NMAE | Generated min | Generated max | Truth min | Truth max |
|---|---|---|---|---|---|---|---|---|
| U (m/s) | 0.224 | 0.141 | 2.299% | 1.442% | -2.338 | 7.324 | -2.435 | 7.318 |
| W (m/s) | 0.123 | 0.077 | 1.884% | 1.174% | -2.74 | 3.391 | -2.835 | 3.702 |
| V (m/s) | 0.144 | 0.09 | 1.726% | 1.08% | -4.095 | 3.897 | -4.118 | 4.207 |
| T (°C) | 0.227 | 0.134 | 1.773% | 1.051% | 28.745 | 40.571 | 27.97 | 40.753 |

Table 1 presents the overall statistic error based on validation metrics discussed in Section 2.3. As opposed to common RMSE errors of mean flow field prediction from previous studies, such as 0.35 m/s (local-FNO using CityFFD dataset), 0.38 m/s (best legacy model), and 2.23 m/s (CNN full urban field), the overall normalized RMSE across the three velocity components is around 1.925%, confirming that the generated ensemble is physically anchored to the conditioning mean flow input without systematic drifting, especially for the dominate wind direction u with RMSE of only 0.224 m/s (Chockalingam et al., 2023; Qin et al., 205a. The temperature channel continues to achieve an exceptional RMSE of 0.227 and NRMSE of 1.773%, which reflects the complex, intermittent nature of heat transportation in plume and often exhibits sharper gradients than the velocity fields. These values are well within engineering tolerances and confirm that the model has successfully learned the underlying coherent flow structures of the urban morphology. The $w$ and $v$ have even smaller normalized RMSE of 0.123 m/s and 0.144 m/s, since the fluctuation of the wind in these two minor directions are significantly smaller than $u$. the reference wind velocity $U_{mean}$ In this case, even the RMSE and MAE is very small for these two variables, they incur a higher normalized value. This disagreement of high normalized errors from low absolute errors exhibits the model's robustness. It demonstrates that the model effectively captures subtle, low-magnitude velocity fluctuations without suffering from numerical instability from low nominators.

## 3.4 Turbulence generation

For wind-flow snapshots, standard deviation is used because it gives a compact, directly measurable estimate of fluctuation amplitude over a sampling window, and many wind-engineering workflows already define turbulence through wind-speed standard deviation or turbulence intensity derived from it (Ren et al., 2018). In Figure 10a, both ensembles is able to identify $u$- and $v$- component fluctuations along major flow paths, around windward building corners, and within downstream wake and recirculation regions. The corresponding difference maps show that most of the errors are concentrated primarily along narrow shear layers and building edges rather than being distributed uniformly across the domain. Such level of agreement is also indicated by the relatively low normalized errors for all variables in Table 2: the $u$ field has an NRMSE of 7.17% and an NMAE of 4.98%, while $v$ achieves the lowest NRMSE and one of the lowest NMAE values, at 6.82% and 4.50%, respectively. The particularly good performance for $v$ is reasonable because of its high similarity of the highly dynamic regions near the central building cluster and the long wake in both the horizontal and vertical sections. For $u$, the slightly higher RMSE of $0.086\ \mathrm{m\ s^{-1}}$, compared with an MAE of $0.055\ \mathrm{m\ s^{-1}}$, reflects several localized red and blue regions in the difference maps, especially around building corners and near the lower-left high-variability zone, where small spatial displacements of wake structures produce relatively large pointwise errors. This is also visible in Figure 10b, in the vertical contours. The $w$-component also shows good overall agreement, with an RMSE of $0.067\ \mathrm{m\ s^{-1}}$and an NRMSE of 7.03%; however, its NMAE of 5.87% is the largest among the velocity components. This is consistent with the more small-scale difference around rooftop shear layers and above the central buildings in the vertical section and the fact that the $w$-component has a lower overall fluctuation magnitude, as highlighted in Figure 10b. For temperature, the generated ensemble successfully reproduces the general pattern of plume movement, the vertical fluctuation along building height and downstream of the buildings, and more stable regions outside the plume. Nevertheless, temperature has the highest NRMSE, 8.84%, and an RMSE of 0.118 ℃ , which correspond to the localized differences along the plume boundaries and immediately above the rooftops. Its relatively low NMAE of 4.21% and MAE of 0.052 ℃, however, indicate that these larger errors are spatially limited and that most of the thermal field is reproduced accurately.

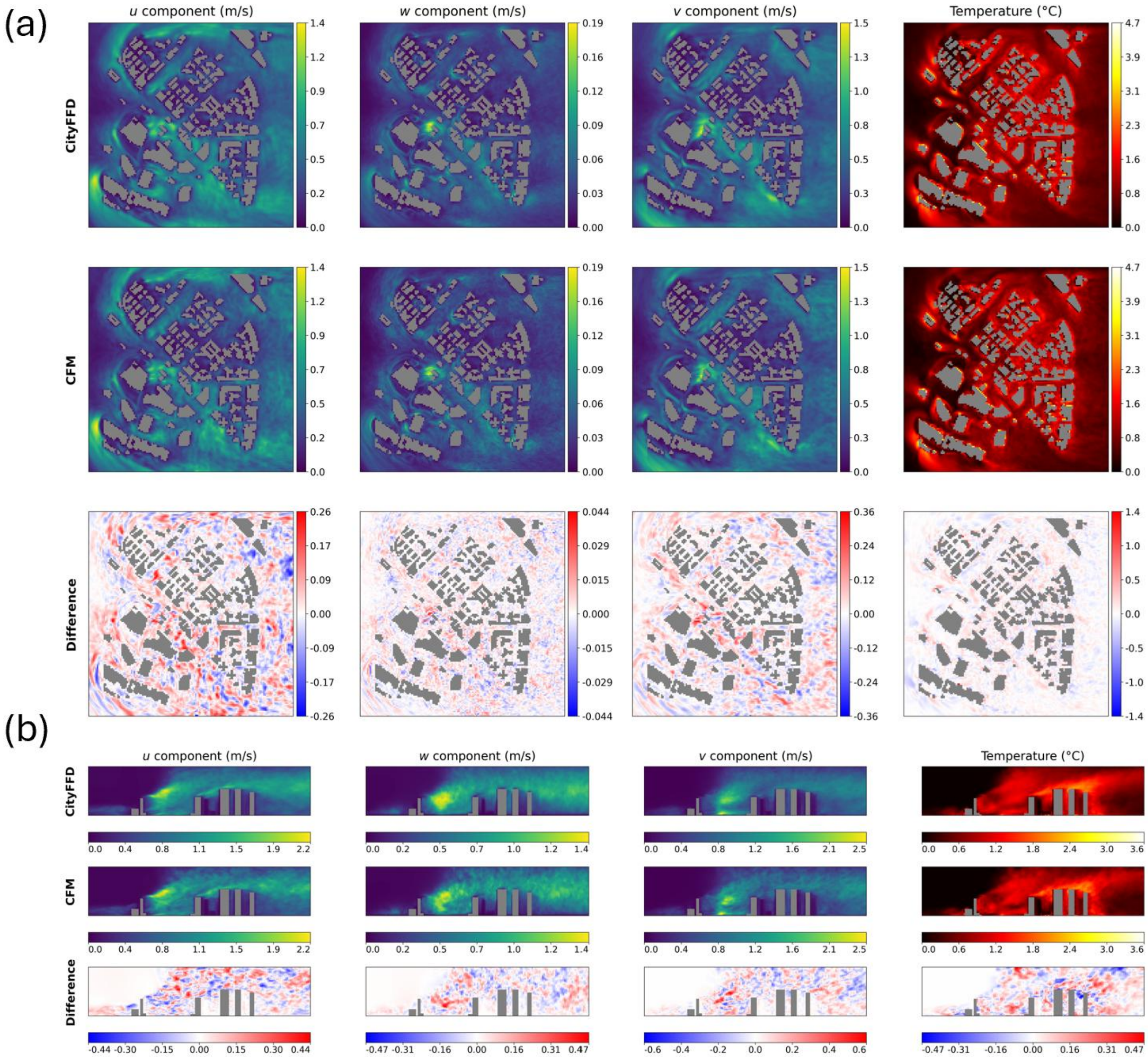

*Figure 10: Standard deviation maps. Top row: generated (35 snapshots); middle row: CityFFD (35 snapshots); bottom row: difference, for (a) z= 3 and (b) central yz plane*

*Table 2 Error matrix for Standard Deviation Map for generated ensemble vs CityFFD reference ensemble (35 snapshots)*

| Variable | RMSE | NRMSE | MAE | NMAE |
|---|---|---|---|---|
| U (m/s) | 0.086 m/s | 7.17% | 0.055 m/s | 4.98% |
| W (m/s) | 0.067 m/s | 7.03% | 0.043 m/s | 5.87% |
| V (m/s) | 0.073 m/s | 6.82% | 0.046 m/s | 4.50% |
| T (°C) | 0.118 (°C) | 8.84% | 0.052 (°C) | 4.21% |

Overall, the consistently low MAE than RMSE when compared against CityFFD standard deviation map, together with the predominantly small differences along building edges and shear-layer boundaries, indicates that the generated ensemble captures the second-order statistics well while having localized errors for some highly turbulent and variable locations.

Across urban-boundary-layer studies, TKE consistently captures production, transport, and dissipation of turbulence, whereas velocity standard deviations describe only fluctuation amplitude and miss how energy is generated, redistributed, or lost (Stull, 1988; Akinlabi et al.,2023). In Figure 11, it can be found that similar spatial energy distributions highlight the model's capability in localized feature extraction. In the horizontal planes (XY), the CFM effectively replicates the high-energy hotspots generated by shear layers along the windward corners of the building clusters around coordinates of (200m, 1000m), where an obvious and within the high-velocity bypass streams. Crucially, the vertical cross-sections (XZ) confirm that the model captures the sharp, discontinuous TKE gradients at the building-fluid interface, accurately reproducing the elevated turbulence generated in the shear layers right above the rooflines $z \approx 120\ m$ rather than offering a blurred or over-smoothed representation.

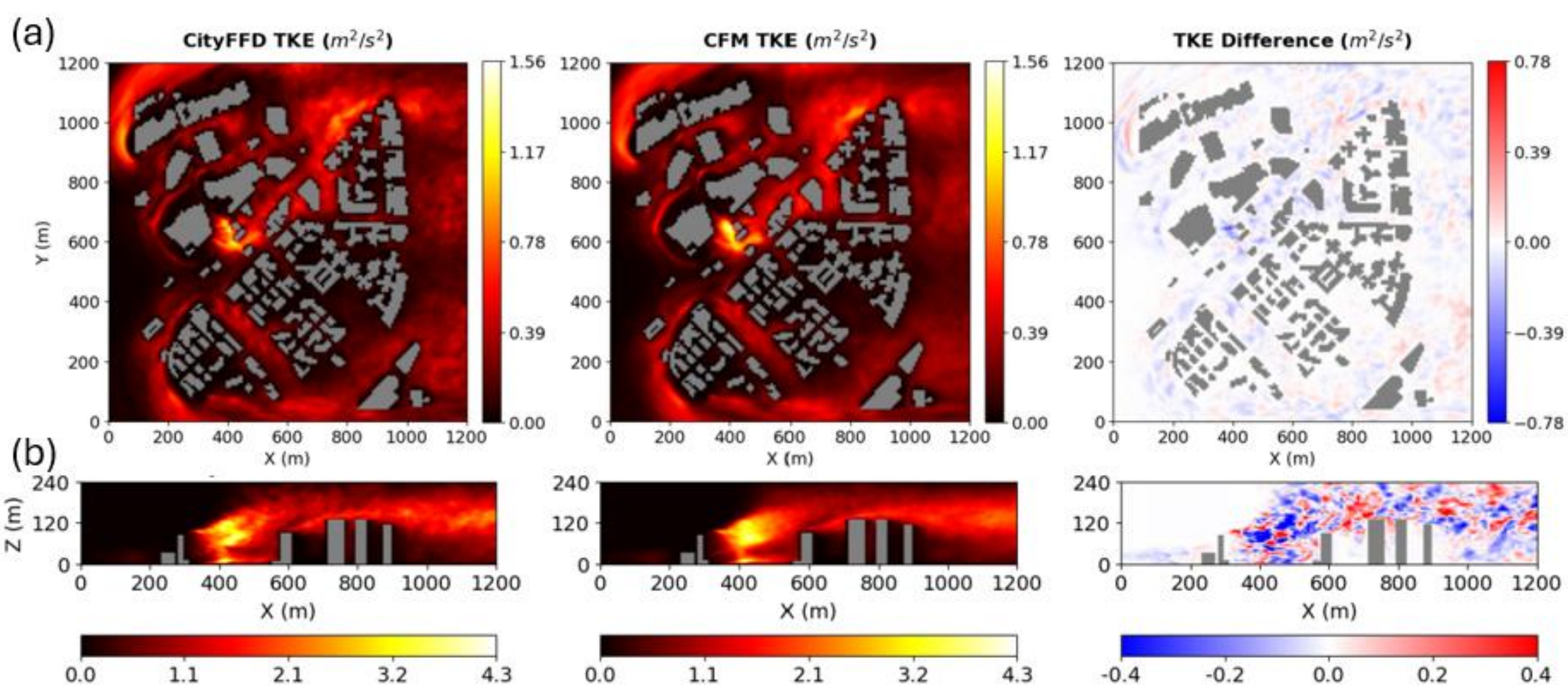

*Figure 11: TKE maps. Left: CityFFD (35 snapshots); Center: CFM (35 snapshots); Right: difference. for (a) z= 3 and (b) central yz plane*

Looking at the TKE Difference maps, the errors are stochastically distributed as balanced, alternating positive (red) and negative (blue) patches in the wake regions. This structural pattern indicates the absence of systemic under- or over-dispersion bias, proving that the generative model successfully preserves both the local structure and total magnitude of turbulent energy cascade within complex urban configurations.

*Table 3 Statistics of Turbulent Kinetic Energy map (35 CFM snapshots vs. 35 CityFFD snapshots)*

| | RMSE ($m^2/s^2$) | MAE ($m^2/s^2$) | NRMSE (refer to mean TKE) | NMAE (refer to mean TKE) | Generated min TKE ($m^2/s^2$) | Generated max TKE ($m^2/s^2$) | CityFFD min TKE ($m^2/s^2$) | CityFFD max TKE ($m^2/s^2$) |
|---|---|---|---|---|---|---|---|---|
| **TKE** | 0.356 | 0.199 | 7.052% | 3.937% | 0 | 4.608 | 0 | 5.049 |

As shown statistically in Table 3, globally, the CFM model demonstrates high precision, yielding a domain-wide Root Mean Squared Error (RMSE) of $0.356\ \mathrm{m}^2/s^2$, a Mean Averaged Error (MAE) of $0.199\ \mathrm{m}^2/s^2$, along with a range-normalized RMSE and MAE of 7.052% and 3.937%. While both models correctly identify the minimum TKE (0 $m^2/s^2$), there is a discrepancy at the maximum TKE magnitude. The CityFFD baseline calculates a maximum TKE of 5.049 $m^2/s^2$, while the CFM generates a maximum of 4.608 $m^2/s^2$ for the same number of snapshots. This smoothing effect is a common characteristic of generative when applied to chaotic fluid systems (Oommen et al., 2026). While the underlying U-Net architecture is highly capable of spatial mapping, the continuous regression objective of flow matching optimizes for the most probable, stable probability paths. As a result, the model tends to regress toward the expected mean states—excellently capturing the bulk flow and larger coherent structures, but slightly dampening the most extreme, chaotic, and high-frequency peaks found in the center of turbulent shear layers.

## 3.5 Probability density function

The alignment of PDFs and KDE curves across multiple heights provides strong evidence of distributional accuracy and its further explored in this subsection. As shown in Figure 12, the CFM output major overlays the CityFFD ground truth across all four components $u, w, v, T$. The model captures the transition from skewed, near-wall distributions at lower elevations to more symmetrical profiles at higher altitudes. The high degree of overlap in the tails of the velocity PDFs confirms that the model accurately samples the rare, high-magnitude turbulent fluctuations that are physically characteristic of urban wind fields. The temperature PDFs at higher elevations exhibit distinct bimodal peaks near 32 °C and 40 °C. The CFM model does not average these peaks into a single distribution; instead, it precisely replicates the sharp intensity of the secondary peak, indicating that the flow-matching objective can learn complex, non-Gaussian scalar transport phenomena where sharp thermal gradients are present. The KDE curves maintain consistent smoothness and bandwidth matching with the ground truth, confirming that the model has avoided mode collapse — a common failure of GAN-based architectures.

The Kernel Density Estimation of PDF analysis across multiple heights $z = 3m, 80m, 160m, 240m$ demonstrates that the CFM model accurately captures the underlying statistical distributions of the urban flow field. For the primary streamwise velocity (u) and temperature (T), the CFM captures complex, multimodal distributions exceptionally well. At $z = 80$m, the model perfectly traces the dual peaks of both u and T, proving it accurately handles spatial intermittency and distinct micro-climate states. However, non-Gaussian behavior in the turbulent vertical (w) and lateral (v) velocity components reveals subtle discrepancies, primarily near the ground $z = 3$ m at this near-surface level, the CFM overestimates the peak probability density of w, failing to completely match the broader, multi-peak structure of the CityFFD reference. This discrepancy indicates that while the CFM aligns excellently at higher elevations, it tends to over-smooth highly localized, terrain-induced boundary layer turbulence near the surface.

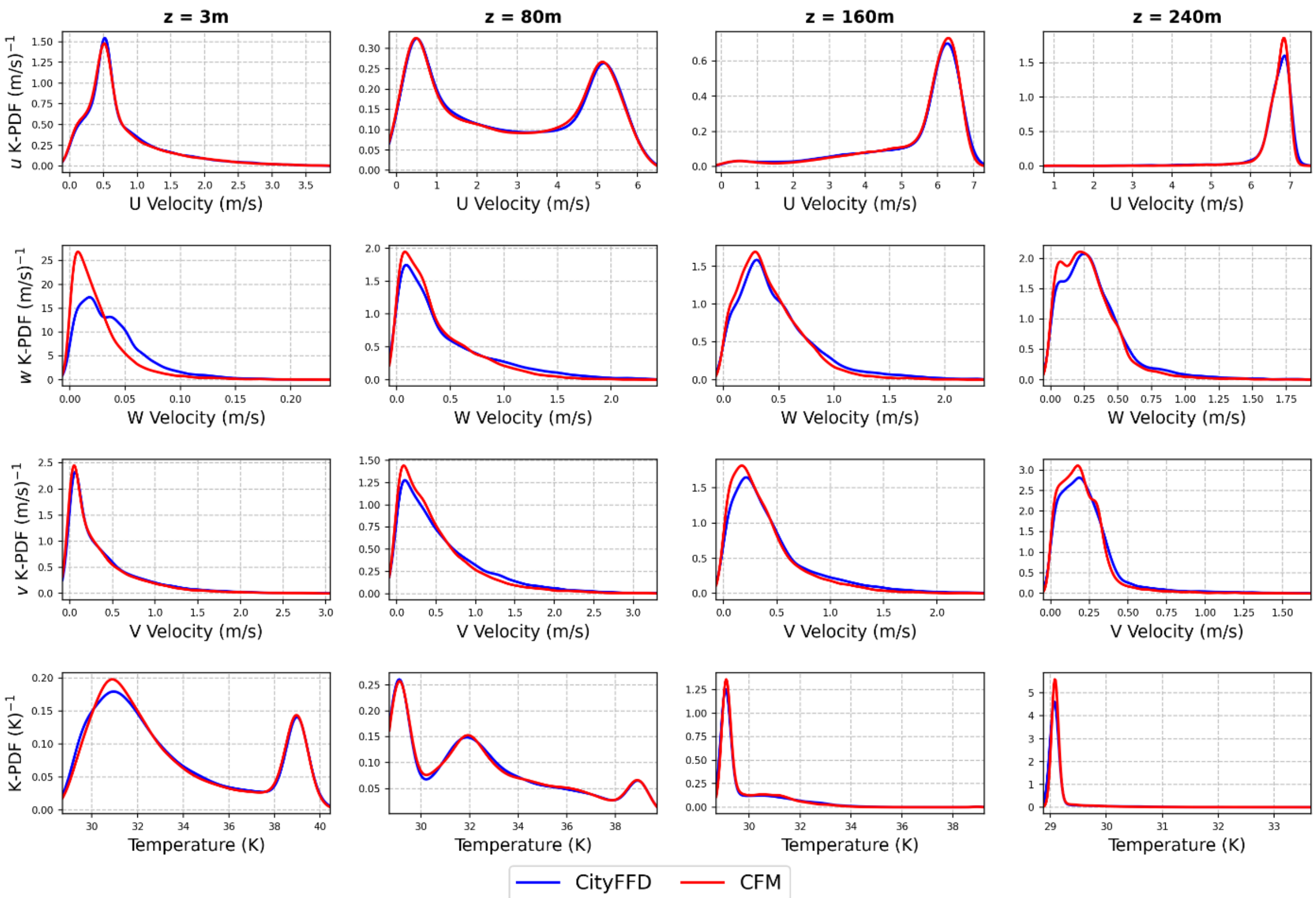

*Figure 12: Kernel-PDF for u, w, v, T at multiple heights. Red: CFM (35 snapshots); blue: CityFFD reference (35 snapshots) for z=3m, 80 m, 160 m, and 240 m*

## 3.6 Vertical wind profiles

To further assess the validity of the CFM results in this subsection the profiles are presented in representative locations of the urban microclimate. As for the wind profile plot, the midpoints of profiles are calculated based on mean of the 35 snapshots and bars are calculated based on the range of different variables from the ensemble as mentioned in section 2.4.

The vertical profile validation for the three characteristic locations is shown in Figure 13, demonstrates that the generative model (CFM) highly accurately captures mean atmospheric structures, though its precision slightly degrades when resolving micro-scale turbulent fluctuations in highly obstructed spaces. Globally, the model excels at predicting primary streamwise velocity (u) and temperature (T) profiles across all urban terrains, successfully reproducing macroscopic boundary layer shear and thermal stratification physics. Further validation metrics for those locations are presented in Table 4. This strength is highlighted at the Upstream site, where the u component perfectly matches the logarithmic profile with an exceptional $r^2$of 0.996 and a low NRMSE of 2.21%.

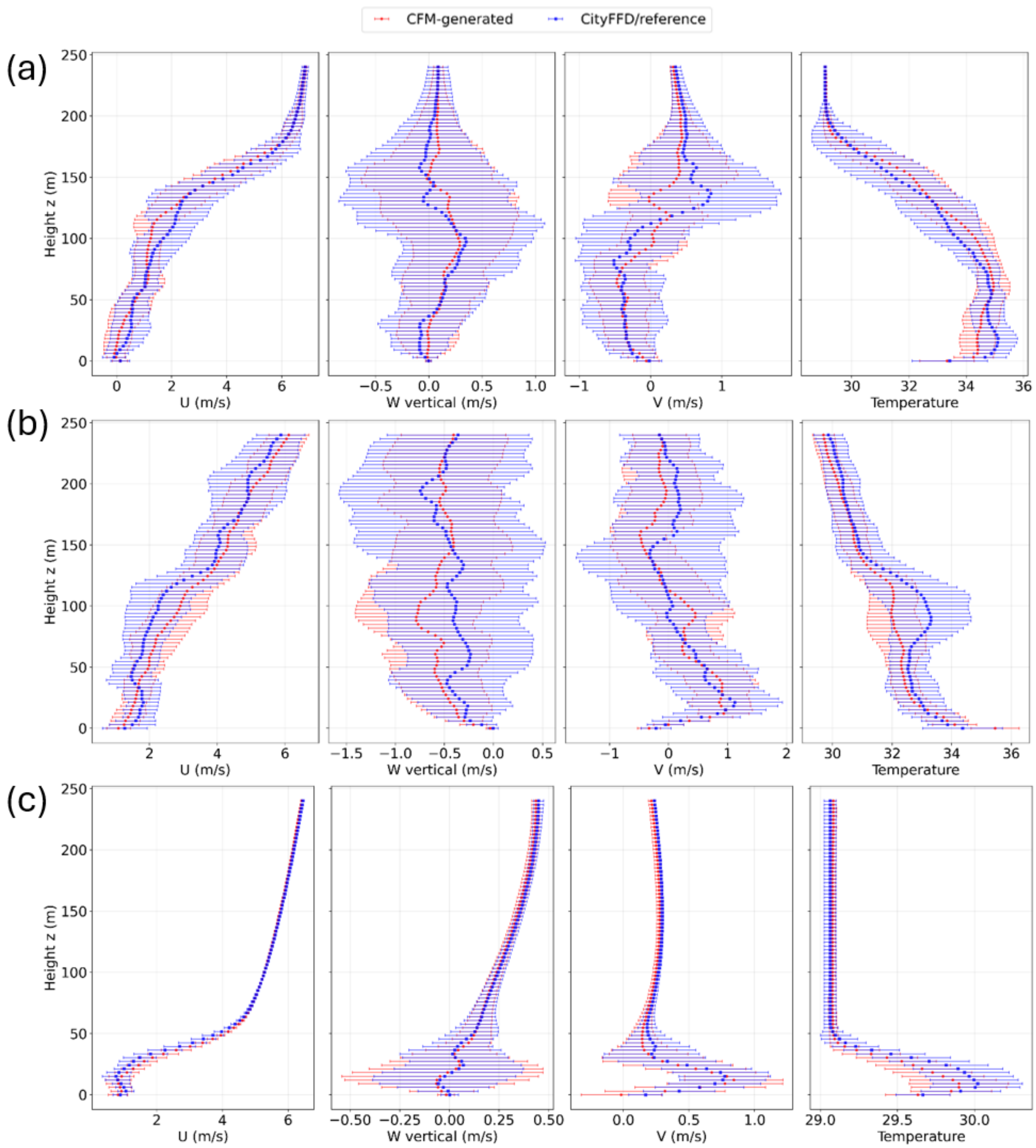

*Figure 13: Multi-site vertical profiles for u, w, v, T at (a) open area, (b) city canyon, and (c) upstream locations (last row). Solid lines: ensemble mean; horizontal bars: variable range at each height for ensemble.*

*Table 4 Validation metrics of vertical profiles in selected locations*

| Profiles | channel | RMSE (m/s or ℃) | NRMSE (%) | $R^2$ |
|---|---|---|---|---|
| Upstream | u | 0.125 | 2.210 | 0.996 |
| | w | 0.013 | 2.594 | 0.995 |
| | v | 0.041 | 6.810 | 0.952 |
| | T | 0.064 | 6.732 | 0.985 |
| City Canyon | u | 0.351 | 7.617 | 0.970 |
| | w | 0.196 | 26.56 | 0.029 |
| | v | 0.228 | 15.83 | 0.72 |
| | T | 0.517 | 11.47 | 0.895 |
| Open Area | u | 0.258 | 3.721 | 0.993 |
| | w | 0.073 | 16.935 | 0.732 |
| | v | 0.239 | 17.628 | 0.732 |
| | T | 0.347 | 5.755 | 0.978 |

Remarkably, the CFM even resolves complex, non-linear "S-shaped" velocity profiles generated by canyon vortices within the Street Canyon. However, localized weaknesses emerge in highly turbulent crossflows; the model severely underestimates ensemble variance and vertical velocity fluctuations (w) within the dense Street Canyon, resulting in an elevated NRMSE of 26.56% and a poor $r^2$ of 0.029. Lateral velocity (v) similarly exhibits higher relative errors in both the canyon (15.83%) and open park (17.63%) zones due to the highly intermittent nature of crosswind wake shedding. From an engineering perspective, this validation proves that while CFM is a highly dependable tool for evaluating pedestrian comfort, general thermal plumes, and steady state mean wind loads, it tends to over-smooth localized, three-dimensional turbulence. Therefore, engineers should exercise caution if relying on this model's localized vertical velocity variance for structural fatigue designs or peak wind load calculations inside deep street canyons.

# 4. Discussion

The results presented in this paper demonstrate that CFM provides an effective generative surrogate for producing ensembles of LES instantaneous urban microclimate fields. Two aspects of the contribution merit further discussion: what the stochastic ensemble enables that no deterministic surrogate can provide, and what the observed limitations reveal about the current model. The fundamental distinction between this approach and existing deterministic surrogates lies in the nature of the output. A deterministic model — whether CNN, FNO, or GNN — produces a single prediction for a given geometry. Even when that prediction is accurate, it cannot support probabilistic analysis. In contrast, the CFM-generated ensemble of 35 independent snapshots enables probabilistic assessments that were previously accessible only through full LES campaigns. For pedestrian wind comfort, the ensemble makes it possible to compute the probability of exceeding a discomfort threshold at each point in the domain, yielding an exceedance probability map rather than a binary pass/fail classification based on the mean wind speed. For natural ventilation assessment, the ensemble provides a distribution of instantaneous pressure differences across building openings rather than a single value, enabling the calculation of ventilation rates with uncertainty bounds. For structural wind loading, the tail behavior of the velocity distribution — which the PDFs in Figure 12 show is accurately captured — determines peak dynamic loads that govern design. These applications require not merely a diverse set of snapshots, but physically and statistically consistent ones, which is precisely what the validation results confirm.

The results also reveal an honest limitation: the model exhibits a slight tendency to act as a low-pass filter, occasionally underestimating peak fluctuation intensities in high-fluctuation zones. This behavior is visible in the blue-dominant regions of the standard deviation difference maps (Figure 10) and in the slight underestimation of peak TKE values (3.86 vs. 3.99 $m^2/s^2$). This is a recognized characteristic of generative models that learn a transport map from noise to data: the averaging inherent in the training objective tends to soften the sharpest extremes. In practice, this effect is small — the standard deviation differences are 0.35 m/s and 0.33 °C — and the tail behavior of the PDFs remains well preserved, but it should be acknowledged for applications where peak gust prediction is critical.

The practicality of patch-based generation strategy with shared noise initialization extends beyond this specific application. The challenge of generating spatially coherent fields from locally trained generative models is common to any 3D application where memory constraints preclude full-domain training. The shared-noise initialization and per-step cosine-weighted merging introduced here maintain both spatial continuity and turbulent energy across patch boundaries — two requirements that are typically concentrated by researchers, since smoothing at boundaries naturally suppresses fluctuations. The visual and quantitative results confirm that this scheme provide a successful alternative.

Furthermore, it is worth noting the computational advantage of CFM. Once trained, the model generates a single 3D instantaneous snapshot in seconds on a single GPU, compared to hours for CityFFD. An ensemble of 35 snapshots requires less than 10 minutes rather than hours. This cost reduction enables the possibility of integrating stochastic microclimate assessment directly into iterative urban design workflows, where parametric studies across building configurations currently rely on RANS or deterministic surrogates that cannot capture the turbulence phenomena central to comfort, ventilation, and safety (Wu and Quan, 2024).

To further validate the model's reliability for urban design, wind comfort, and in order to bridge the scientific outputs to potential real-world applications, we evaluate its performance in a critical engineering parameter; the wind gust. Local wind extremes are a critical consideration in environmental wind engineering, but also it is closely related to structural applications. Here, we focus on environmental aspects, since wind induced pressures are not part of the targets of this work. Although, an initial understanding of the capabilities of CFM to generate appropriate gusts can ne revealing into estimating the readiness of this model for structural application in the future. The gusts were calculated based on the quasi-steady assumption around a region of three characteristic targeted buildings in the urban area. Davenport's classic Peak Factor Method was used (Davenport, 1967), which is a standard wind engineering methodology for designing and optimizing against peak wind scenarios, namely under $X_{peak} = \mu + g\sigma$, where the peak wind velocity is defined by mean and standard deviation of the instantaneous flow field. Under an approximate Gaussian assumption, $g = 3$ corresponds to a three-sigma upper estimate. Although local wind speed distributions may deviate from normality, this metric provides a consistent quasi-steady indicator for comparing CFD and generated ensembles.

As indicated in Figure 2b, three characteristic areas were selected to represent distinct urban morphological conditions and their associated local wind-flow mechanisms. The first area is centred on the tallest building, where strong exposure, flow separation, and corner acceleration can produce elevated mean velocities and wind-speed fluctuations. The second area contains a relatively isolated building, providing a comparatively unobstructed setting in which building-induced acceleration, wake formation, and downstream flow recovery can be evaluated. The third area represents an urban-canyon configuration, where interactions among closely spaced buildings may cause sheltering, flow channeling, recirculation, and strong spatial velocity gradients. For each location, a 200 m x 200 m x 240 m volume was defined around each target building to include both the immediate building-scale response and the surrounding wake or interference region. This is also depicted in Figure 2b, where its of those regions are displayed, with corresponding arrow or the vertical planes that were used for comparisons. Together, these areas encompass contrasting high-exposure, isolated-wake, and dense-building-interference conditions to assess the model's ability to reproduce localized quasi-steady wind extremes. Results for the vertical contours are presented in Figure 13, for the horizonal contours at z = 3 m in Figure 14 and finally the data in the entire volume of each region are correlated in Figure 15.

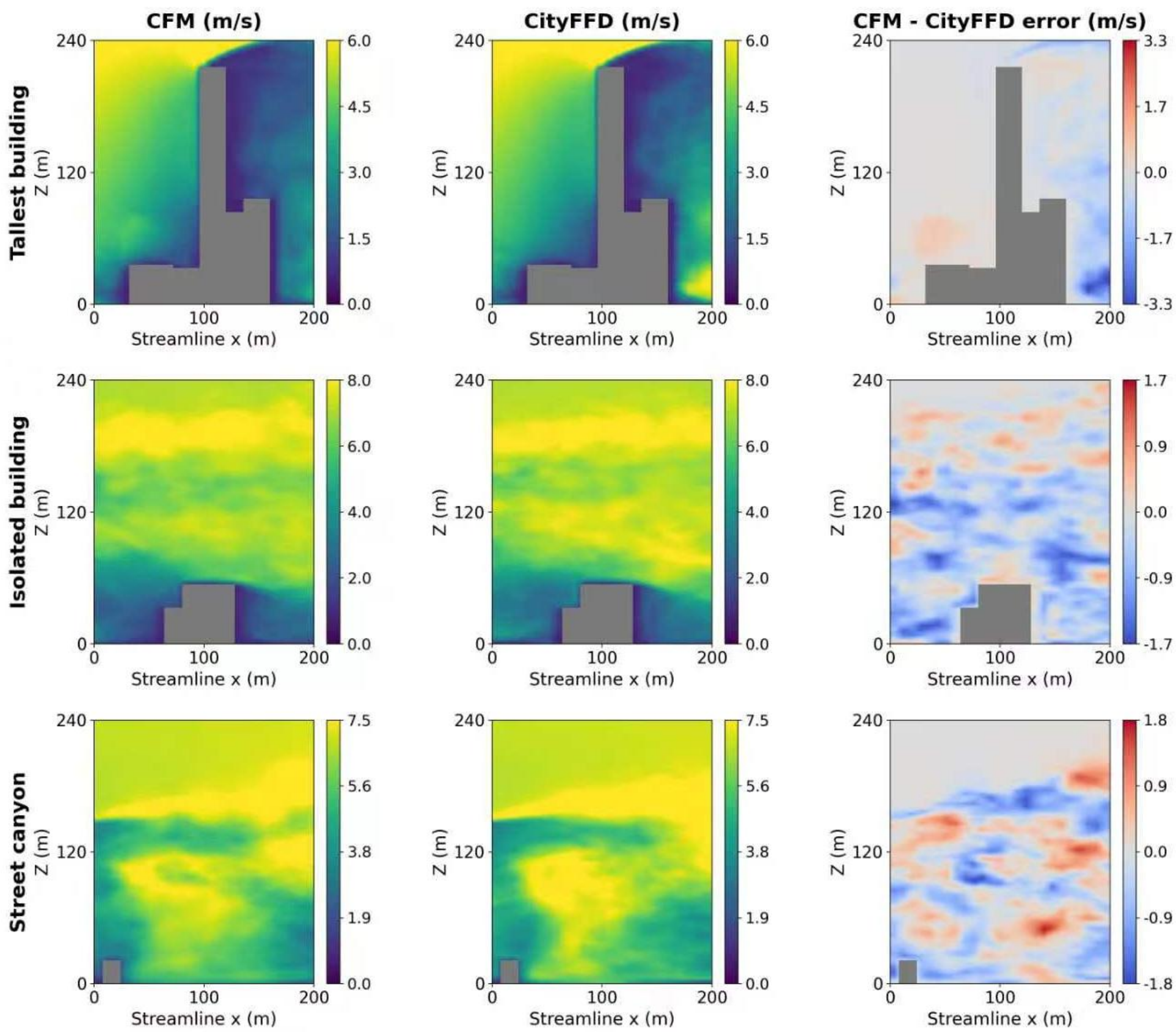

*Figure 14: Wind gusts in vertical planes for regions of tallest building (top), isolated building (middle), and street canyon (bottom) from CFM, CityFFD and their error*

First, regarding the vertical contours in Figure 14 and the tallest building, CFM captures the upstream flow and wake recovery of the gusts with good accuracy. As seen in the first row of Figure 14, the size of the gusts immediate before, above and after the building are very similar, while the magnitude of the gusts have a great match with the CityFFD values. Their differences that are displayed in the third column are mainly below 0.5 m/s, while some larger discrepancies are found downstream and in lower heights. The isolated building cases presented in the second row of Figure 14, displays a similar accuracy although in this case the gusts have a totally different formation. Largest gusts are seen above the urban canopy at heights above ~60 meters, since the flow is less obstructed at this range. Even at these large heights, with gusts reaching the maximum incoming velocity of the reference CityFFD data, CFM captures complex formation of the gusts with good spatial resolution, with errors remain bellow 0.4 m/s for most of the domain, while locally reach ~1 m/s. Finally, in the third row the vertical profile in the street canyon is seen, in which case there are not building strongly obstructing the flow. The wake of the buildings though is seen around the height of 130 m, were the gusts take this peculiar shape. Even in this condition, CFM is capable of capturing the shape of the gusts, with errors remaining mainly below 0.6 m/s.

In Figure 15, the equivalent horizontal figures are presented at 3 m height. Regarding the tallest building in the top row, CFM captures the general form of the gusts around it, but local errors at the lee ward side reach up to 1.5 m/s locally, with CFM underestimating the magnitude of the gusts. The rest of the wake and upwind region are captured with very good accuracy. For the isolated building (Figure 15 second row), CFM predictions are generally in good agreement, except the southeast face, where CFM overpredicts slightly the magnitude of the gusts. The street canyon flow presented at the third row, again proves the ability of CFM to capture gusts with unique formation and curvature, while the magnitude is slightly underestimated locally.

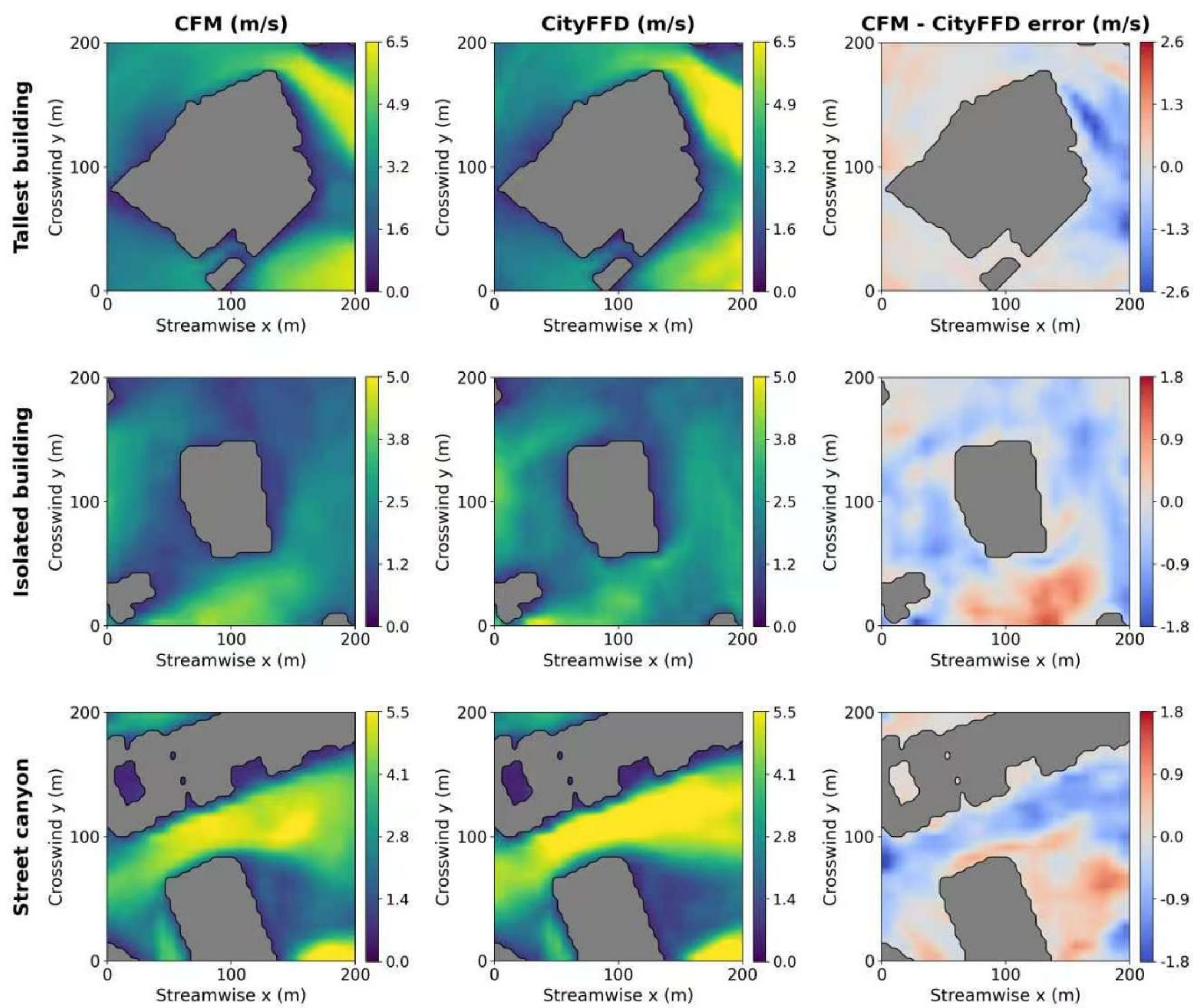

*Figure 15: Wind gusts in horizonal planes (z = 3m) for regions of tallest building (top), isolated building (middle), and street canyon (bottom) from CFM, CityFFD and their error*

In order to better understand the discrepancies seen in Figures 14 and 15, in Figure 16 the correlation plots between the cell by cells comparison of CFM and CityFFD for those three characteristic volumes are presented. In each subfigure four statistical metrics are included; RMSE, MAE, bias and r. As seen, the overall accuracy of MAE is below 0.3 m/s for the tallest building and the street canyon case, while it is slightly increased for the isolated building. Liner regions metrics are about 0.98 for the two, with the isolated case again presented a slight decrease to 0.87. The phenomenally simpler volume of the isolated case is actually more difficult to capture since in this location turbulence features are advected from upstream, thus turbulence is not generated solemnly due to the configuration of the building, which makes it harder to predict for CFM. Also, in this region the gusts have smaller magnitude compared to the other two cases, which complicates the prediction process even further. Still, even in this extreme case, the accuracy is CFM is excellent overall and in line with the targets of this work.

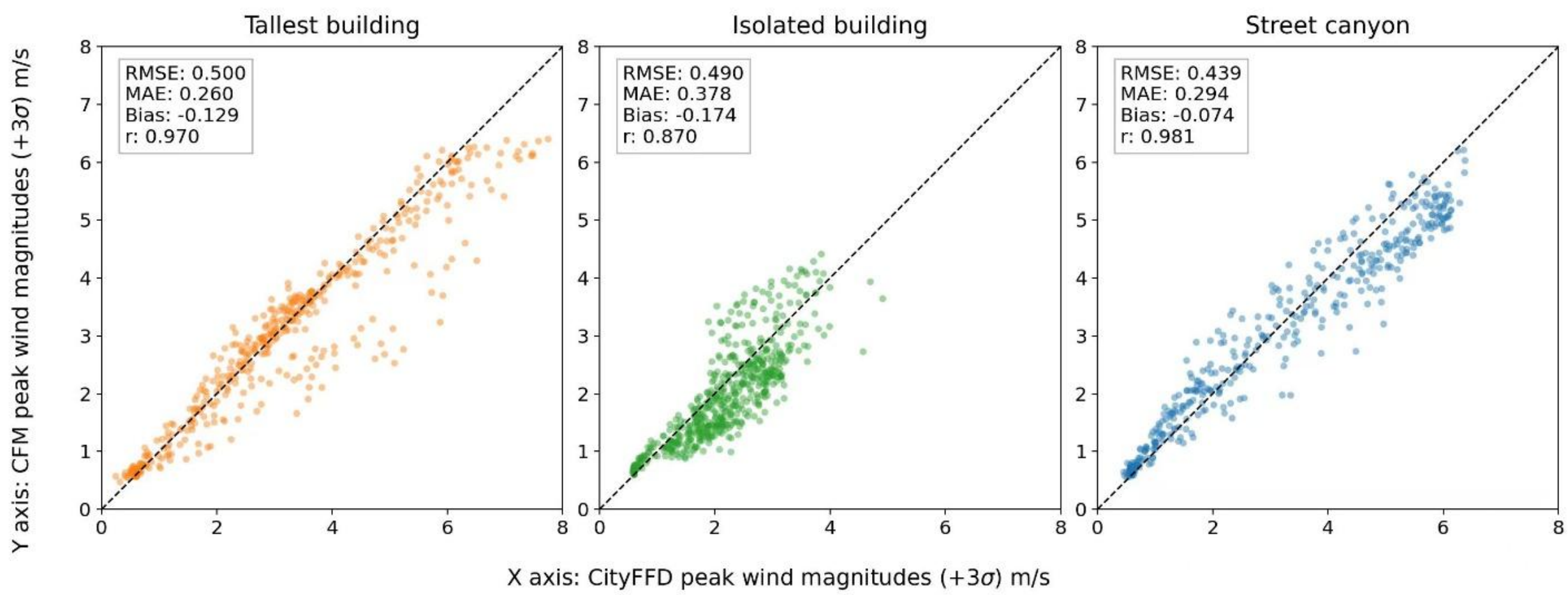

*Figure 16: Parity plot of CFM predicted and CFD-simulated local 3σ peak wind speeds across 3 different urban configurations (Tallest building, Isolated building, and Street canyon)*

Building upon the established framework of CFM, future research should prioritize the transition from static snapshot synthesis to the generation of continuous spatiotemporal wind flow series. By integrating recurrent mechanisms or temporal attention layers, the model should evolve to reproduce the coherent evolution of turbulent structures, representing time-dependent phenomena such as peak gust impacts and transient pollutant transport. To enhance the model's adaptability to diverse urban morphologies, the adoption of Graph Neural Networks (GNNs) for geometric embedding represents a critical next step. Unlike rigid pixel-based grids, GNNs can treat urban environments as unstructured graphs where buildings serve as nodes and airflow paths as edges, allowing the model to generalize across complex, non-orthogonal city layouts with high spatial efficiency. Furthermore, the input dimensionality should be expanded to include more dynamic multi-variable conditioning, enabling users to modify the wind direction $\theta$ and reference velocity $U_{ref}$ as continuous parameters during inference. This flexibility would transform the generator into a robust testing suite for "what-if" scenarios, ultimately facilitating the integration of high-fidelity wind data into real-time digital twins and responsive urban design platforms that can adapt to shifting atmospheric conditions in seconds. There is still a need to apply the CFM model to more CFD dataset such as OpenFoam or Fluent to see how the workflow can be integrated and bring more impact. Finally, the initial results regarding the gusts present in Section 4, provide confidence to try CFM for structural engineering applications as well.

# 5. Summary - Conclusion

This paper presents a novel application of CFM to the generation of 3D instantaneous urban microclimate fields in complex and realistic urban configurations. The key contributions are as follows:

- The CFM framework, trained on 18 urban configurations with diverse morphologies, conditioned solely on building geometry and mean flow information without any turbulence statistics as input, could generate physically plausible instantaneous velocity and temperature snapshots that reproduce the stochastic character of LES data.
- The patch-based 3D generation strategy, using overlapping $64^3$ voxel patches with shared-noise initialization and variance-preserving weighted averaging, enables full-domain reconstruction of the large 1.2 km x 1.2 km area, otherwise non-feasible due to GPU memory restrictions, while maintaining spatial continuity across the domain. Generation of statistically converges snapshot takes around 6 minutes (10 seconds for each snapshot), making CFM appropriate for iterative design.
- Validation against CityFFD reference data for the tested case were thoroughly reported and confirms that the generated ensembles reproduce the mean flow to within 2.99% NRMSE for wind and 1.77% for temperature. The match to the reference standard deviation was 7.17% NRMSE for wind and 8.84% for temperature, and CFM accurately captured TKE distributions with NRMSE of 7%. Probability density functions including non-Gaussian tails and bimodal distribution were captured with good accuracy as well. Furthermore, local vertical wind profiles were presented at representative urban locations with $R^2$ values exceeding 0.99 for the primary flow component.
- Wind gusts predictions by CFM, based on the quasi-steady assumption, were also reported in three district regions of the urban domain and correlated very well with CityFFD data. Gust sizes and magnitude can be detrimental for pedestrian winds, pollutant and heat transport and a discussion was included as to how to bridge the gap between the scientific findings of this paper and real-world wind engineering applications.

These results demonstrate that CFM provides an effective and computationally efficient path to LES resolution stochastic urban microclimate fields, enabling probabilistic wind and temperature comfort assessment, and wind gust estimation at a fraction of the cost of full LES campaigns.

# Acknowledgments

This research was supported by the Canada First Research Excellence Fund (Volt-Age) under the SEED project, *Creating Electrified and Decarbonized Healthy Urban Microclimate around Building Clusters through Climate-Resilient Solutions*, and under the IMPACT project *Transforming Built and Urban Microclimates: Advancing Resilience Science for Vulnerable Populations in a Decarbonized and Electrified Canada*. The authors would like to thank the guest editors for the invitation to the special issue of Advanced Measurement and Modeling Techniques for Urban Wind Environment. Numerical simulations were performed using Digital Research Alliance of Canada computing resources.